%% file: main.tex
\documentclass[%
 reprint,
 amsmath,amssymb,
 aps,
 pra,
longbibliography
]{revtex4-2}
\usepackage{graphicx}

\usepackage{mathtools}%
\usepackage{placeins}
\usepackage{float}
\usepackage[table]{xcolor}
\PassOptionsToPackage{breaklinks}{hyperref}
\usepackage{xurl} 
\usepackage{braket}
\usepackage{lmodern,CJKutf8,amsmath,amssymb,mathtools,bm,physics,booktabs,url,multirow,xcolor,siunitx}
\usepackage{algpseudocodex}
\usepackage{algorithm}
\usepackage{subcaption}
\usepackage{overpic}
\usepackage{caption}
\usepackage[
    setpagesize=false,
    linktocpage=true,
    bookmarks=true,
    bookmarkstype=toc=true,
    bookmarksnumbered=true,
    bookmarksopen=true,
    colorlinks=true,
    linkcolor=blue,
    citecolor=blue,
    urlcolor=blue,
]{hyperref}

\newcommand{\reffig}[1]{Fig.~\ref{#1}}

\newcommand{\refsec}[1]{Sec.~\ref{#1}}
\newcommand{\refalg}[1]{Alg.~\ref{#1}}

\begin{document}

\preprint{APS/123-QED}

\title{Bayesian inference of general noise-model parameters from the syndrome statistics of surface codes}

\author{Takumi Kobori}%
\email{takumi.kobori@phys.s.u-tokyo.ac.jp}
\affiliation{Department of Physics, The University of Tokyo, Tokyo 113-0033, Japan}
\author{Synge Todo}%
\email{wisteria@phys.s.u-tokyo.ac.jp}
\affiliation{Department of Physics, The University of Tokyo, Tokyo 113-0033, Japan}
\affiliation{Institute for Physics of Intelligence, The University of Tokyo, Tokyo 113-0033, Japan}
\affiliation{Institute for Solid State Physics, The University of Tokyo, Kashiwa, 277-8581, Japan}

\date{\today}
\input{abst}
\maketitle

\input{intro}
\input{bg}
\input{qne}

\input{result}

\input{conclusion}
\input{ack}

\bibliography{bibliography}
\end{document}

%% file: abst.tex
\begin{abstract}
    The performance of error correction in the surface code can be enhanced by leveraging the knowledge of the noise model for physical qubits.
    To provide accurate noise information to the decoder in parallel with quantum computation, an adaptive estimation of the noise model based on syndrome measurement statistics is an effective approach.
    While noise model estimation based on syndrome measurement statistics is well-established for Pauli noise, it remains unexplored for more complex and realistic scenarios such as amplitude damping which cannot be represented as a Pauli channel.
    In this paper, we propose Bayesian inference methods for general noise models, integrating a tensor network simulator of surface code, which can efficiently simulate various noise models, with Monte Carlo sampling techniques.
    For stationary noise, we propose a method based on the Markov chain Monte Carlo.
    For time-varying noise, which is a more realistic scenario, we introduce another method based on the sequential Monte Carlo.
    We present numerical results of applying our proposed methods to various noise models, such as static, time-varying, and nonuniform cases, and evaluate their performance in detail.
\end{abstract}

%% file: intro.tex
\section{Introduction}\label{sec:intro}
Quantum computers offer a new computational paradigm capable of solving certain problems faster than classical computers, with the potential to revolutionize a wide range of fields~\cite{Feynman1982,QCQI,shor1999polynomial,harrow2009quantum,low2017optimal}. 
However, achieving practical quantum computation requires overcoming the challenges posed by noise in physical qubits, which are so fragile that executing stable calculations is difficult.
Quantum error correction (QEC) codes have therefore been proposed to address these challenges and enable fault-tolerant quantum computation (FTQC)~\cite{Steane1996}.
Among various QEC codes, the surface code~\cite{Kitaev2003,Bravyi1998,Dennis2002} is considered one of the most promising candidates for future FTQC\@.
Owing to its favorable geometrical properties, which enable relatively high-threshold and efficient implementation on various quantum computing platforms, it has attracted significant research interest, leading in experimental demonstrations using superconducting~\cite{Krinner2022, Google2023, Google2024} and neutral-atom qubits~\cite{Bluvstein_2023}.

Several proposals have been made to enhance its performance by optimizing decoding algorithms by utilizing noise model information~\cite{Spitz2018, Huang2020, Chubb2021, Darmawan2018,sivak2024optimization,Hockings2025improving}.
Additionally, the grid shape and Clifford deformation of its stabilizer generators have been shown to influence the code's performance and can also be optimized~\cite{Bonilla2021, David2018, Dua2023, Darmawan2017}.
Such noise model knowledge can be prepared in advance by performing quantum process tomography, where the quantum computer has to be operated aside from the main calculations~\cite{Chuang1997, Evans2022}.
In addition to such an efficiency problem, a problem arises in the case of time-varying noise models~\cite{EtxezarretaMartinez2021,Kilmov2018fluctuations}.
When the noise model changes over time, it becomes necessary to update the noise model information through additional quantum computations for tomography of quantum noise channels.
Estimating the noise channel in parallel with syndrome statistics during decoding effectively addresses these issues.
This approach eliminates the need for a demanding prior tomography procedure and enables online learning of time-varying noise model parameters with no extra quantum overhead apart from the syndrome measurements inherent to the QEC procedure.

Although there have been some attempts at the noise model estimation based on syndrome statistics, only the Pauli noise model case has been considered so far and well-studied~\cite{Wagner2021,Wagner2022,Hockings2025scalable,Spitz2018, Huo2017, Kelly2016}.
There are several reasons why Pauli noise estimation is being actively studied.
First, both Pauli noise and stabilizer code simulations can be efficiently performed using stabilizer methods, which work in polynomial time with respect to the number of qubits on classical computers~\cite{Aaronson2004, Gidney2021_stim}.
Second, the initial logical information does not affect the syndrome statistics in the Pauli noise model, so the estimation can be performed without considering the initial logical information.
Third, the general noise model can be projected to the Pauli noise by randomized compiling with additional overhead~\cite{Wallman2016}.
Although many successful methods exist for estimating Pauli noise models, extending estimation to more realistic noise models beyond Pauli models can be valuable as it can mitigate the need of randomized compiling, which requires additional circuit executions or classical overhead, and can enhance decoder performance through more detailed noise model information.
Such general noise model estimation can be realized by Bayesian inference based on syndrome statistics, which is the main focus of this paper.

In this paper, we propose Bayesian inference method for noise model estimation based on syndrome statistics by integrating tensor network (TN) simulator of the surface code with Monte Carlo sampling methods.
Bayesian inference can offer a more general framework for noise model estimation, allowing us to estimate noise models that cannot be represented as Pauli channels.
For applying Bayesian inference to noise model estimation, we need to consider two points: (1) how to simulate and calculate the posterior distribution of the noise model parameters based on syndrome statistics which cannot be represented analytically and (2) how to calculate the likelihood function of the noise model based on the syndrome statistics.

For the first point, we can use the Monte Carlo sampling methods to generate samples according to the posterior distribution, such as Markov chain Monte Carlo (MCMC) and sequential Monte Carlo (SMC)~\cite{bayes1763lii,Nicholas1953,chib1995understanding,hastings1970monte,Kitagawa1993,Doucet2001}.
MCMC is suitable for datasets prepared before the simulation and SMC is suitable for sequentially generated data of Bayesian inference.
It means that MCMC is suitable for stationary noise case, where the noise parameters are constant and do not change, and SMC is suitable for time-varying noise case, which is a more realistic situation.
For the second point, we focus on the tensor network (TN) simulation~\cite{Orus2019}, which enables efficient representation of noise models and simulation of density matrices for quantum states.
Such TN methods can simulate the surface code with a general noise model in $O(\# qubits)$ by introducing the low-rank tensor approximation efficiently~\cite{Darmawan2017}.
Moreover, we can also calculate the likelihood function based on the syndrome statistics in the TN formulation.
By integrating Monte Carlo sampling methods with the efficient likelihood calculation provided by TN simulation in Bayesian inference framework, effective and versatile general noise model estimation can be achieved.
This approach is applicable to any noise model ideally, including phenomenological and circuit-level noise models, as long as they can be simulated using the TN method.

In these numerical experiments, we report the results of several noise models beyond the Pauli noise models within the range of code-capacity noise models.
We find that the estimation for many cases, such as the amplitude damping, works quite successfully, but for some cases, such as the generalized amplitude damping noise model, it does not.
We also demonstrate that applying the results of the noise model estimation to the TN decoder improves its performance.

The rest of the present paper is organized as follows:
In Sec.~\ref{sec:background}, we briefly review the stabilizer code, surface code and a TN simulation of the surface code.
In Sec.~\ref{sec:qne}, we explain how to estimate the noise model parameters using Bayesian inference and our proposed method based on the TN simulation and the Monte Carlo methods, and discuss the utility of the proposed methods compared to the previous works.
In Sec.~\ref{sec:result}, we show the numerical results of the proposed methods and discuss their efficiency and estimability.
In Sec.~\ref{sec:conclusion}, we summarize the present paper and future issues.

\begin{figure}[t]
    \centering
    \includegraphics[width=0.46\textwidth]{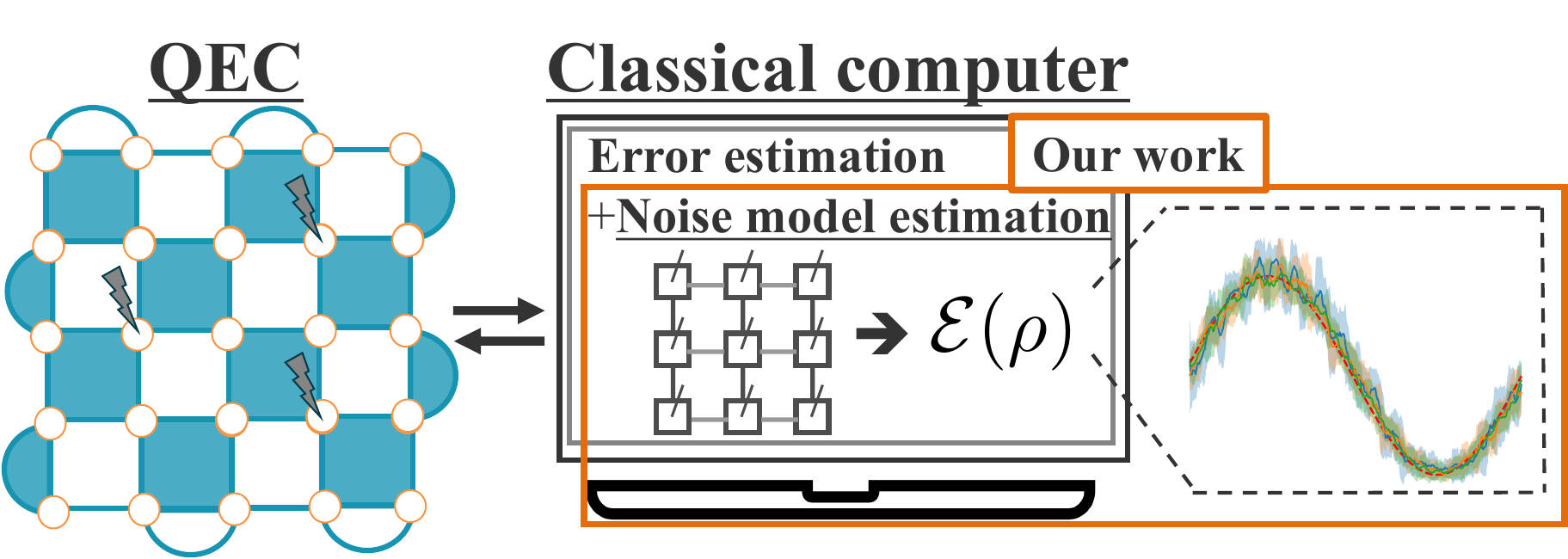}
    \caption{The conceptual figure of key ideas in our work. We propose a novel noise model estimation method based on syndrome measurement results using the TN and Monte Carlo methods. The proposed method can be applied to a broader range of noise models than just the Pauli noise models without any additional quantum overhead.}
    \label{fig:research_position}
\end{figure}

%% file: bg.tex
\section{Preliminary and Tensor network simulation of surface code}\label{sec:background}
\subsection{Stabilizer code}\label{subsec:stabilizer code}
The stabilizer code~\cite{gottesman1997stabilizer} is characterized by a stabilizer group and its generators called stabilizer generators.
The stabilizer group $\mathcal{S}$ is an abelian subgroup of the $n$-qubit Pauli group $\mathcal{P}_{n}$:
\begin{align}
    \begin{split}
        \mathcal{S}= & \{S_i\}\in \mathcal{P}_{n}                                                                  \\
                     & \text{s.t. $-I \notin \mathcal{S}$ and $\forall S_{i},S_{j}\in\mathcal{S}$, $[S_i,S_j]=0$}.
    \end{split}
\end{align}
Each element in $\mathcal{S}\setminus\{ I\}$ has both $\pm 1$ eigenvalues.
According to the eigenvalues of a set of stabilizer generators $\mathcal{S}_{g}$, we can label the Hilbert space as $\mathcal{H}=\bigoplus_{\bm{e}} \mathcal{H}_{\bm{e}}$, where $\bm{e}$ is the set of eigenvalues for the stabilizer generators.
The subspace $\mathcal{H}_{\bm{1}}$ where all eigenvalues are $+1$ is called the code space and is denoted by $\mathcal{C}$.
States within the code space, called code states, are used to encode the original quantum information in the stabilizer code framework.
Because all stabilizers commute, the code state is a $+1$ eigenstate not only of the generators $\mathcal{S}_g$, but of every element of the stabilizer group $\mathcal{S}$:
\begin{align}
    \ket{\psi} \in \mathcal{C} &\iff \forall g_i \in \mathcal{S}_{g},\ g_i\ket{\psi} = \ket{\psi}\\
    &\iff \forall S_{i}\in \mathcal{S},\ S_i\ket{\psi} = \ket{\psi}.\notag
\end{align}
For an $n$-qubit stabilizer code with $|\mathcal{S}_g| = n-k$ independent generators, the code space $\mathcal{C}$ has dimension $2^k$, corresponding to $k$ logical qubits.
Thus, code states can represent arbitrary $k$-qubit quantum information, encoded within the stabilizer code.
\begin{figure}[tbp]
    \centering
    \includegraphics[width=0.4\hsize]{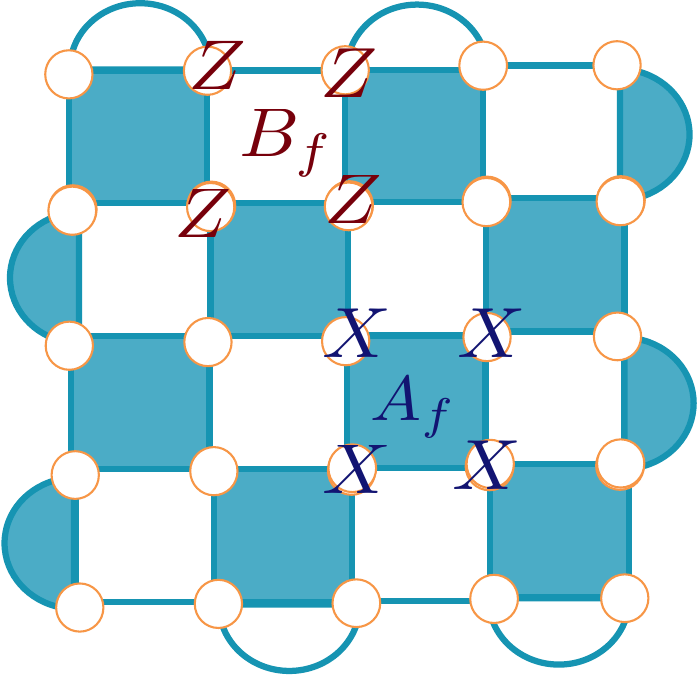}
    \caption{
        Rotated surface code with $d=5$. The number of physical qubits and stabilizer generators is 25 and 24, respectively. This figure only shows the physical qubits but not the qubits for syndrome measurements.
    }\label{fig:rotated}
\end{figure}

Such a redundant representation of quantum information enables protection against noise by monitoring the eigenvalues of the stabilizer generators.
When noise affects the quantum state, it may leave the code space and become a superposition or mixture involving other subspaces $\mathcal{H}_{\bm{e}}$.
Syndrome measurements project the quantum state onto the eigenspace of the stabilizer generators, with the measurement outcomes corresponding to the eigenvalues.
The presence of any $-1$ eigenvalues in the syndrome outcomes indicates the occurrence of errors.
After detecting errors, a decoding algorithm implemented on classical computers can be used to estimate the most likely error locations based on the syndrome measurement results.
Subsequently, errors can be corrected by applying the appropriate Pauli operations to the physical qubits or by updating the Pauli frame~\cite{riesebos2017pauli}.
\subsection{Surface code}\label{subsec:surface code}
The surface code, a type of stabilizer code, is a leading candidate for QEC in future FTQC due to its high threshold and efficient implementation on two-dimensional quantum computing platforms~\cite{Kitaev2003,Bombin2007}.
One type of surface code is represented like \reffig{fig:rotated}.
Its stabilizer generators can be written by only the product of $X$ or $Z$.
In \reffig{fig:rotated}, the blue area represents the product of $X$, and the white area represents the product of $Z$.
Let the lattice be $L\times W$, and its code distance is $d=\min(L, W)$, which means the code can detect all errors of weight $d-1$.
The number of qubits is $LW$ to represent the code and $LW-1$ for ancilla qubits used in syndrome measurements.
Their stabilizer generators are represented by the product of two or four Pauli operators, and the physical qubit on which they act is spatially localized.
Consequently, the surface code can be implemented with qubits arranged in a two-dimensional grid, where operations are performed on adjacent qubits.

\subsection{Tensor network simulation of surface code}\label{subsec:TN simulator}
The TN methods have emerged as one of the most powerful and versatile numerical tools for the simulation of quantum many-body systems.
The central idea of TNs is to efficiently compress the exponentially large Hilbert space by representing quantum states as networks of interconnected tensors, which are multidimensional arrays.
In this framework, each tensor is depicted as a node with legs, where each leg corresponds to an index, and the size of the leg, called the bond dimension, determines the amount of elements stored in the tensor.
Connection of the legs represents the contraction of the indices of the tensors, which is the process of summing over the indices of the tensors.
By using the representative class of TNs, projected entangled pair states (PEPS)~\cite{Verstraete2004} and matrix product states with matrix product operators (MPS-MPO)~\cite{Schollw_ck2011}, can be efficiently simulated with resources that scale polynomially with the number of qubits~\cite{Darmawan2017}.

\begin{figure}[tbp]
    \begin{minipage}[b]{0.48\columnwidth}
        \begin{overpic}[keepaspectratio,width=0.8\columnwidth]{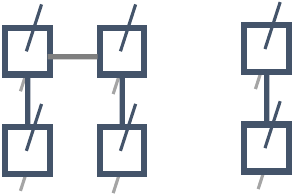}
        \end{overpic}
        \put(-110,60){{\textsf{(a)}}}
        \vspace{0.01cm}
        \begin{overpic}[keepaspectratio,width=1.\columnwidth]{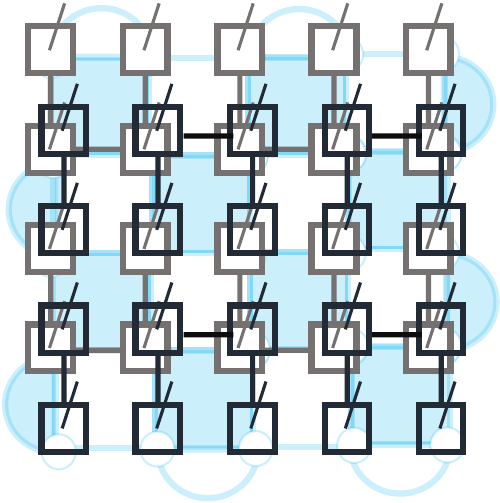}
        \end{overpic}
        \put(-120,120){{\textsf{(b)}}}
    \end{minipage}
    \begin{minipage}[b]{0.48\columnwidth}
        \begin{overpic}[keepaspectratio,width=\hsize]{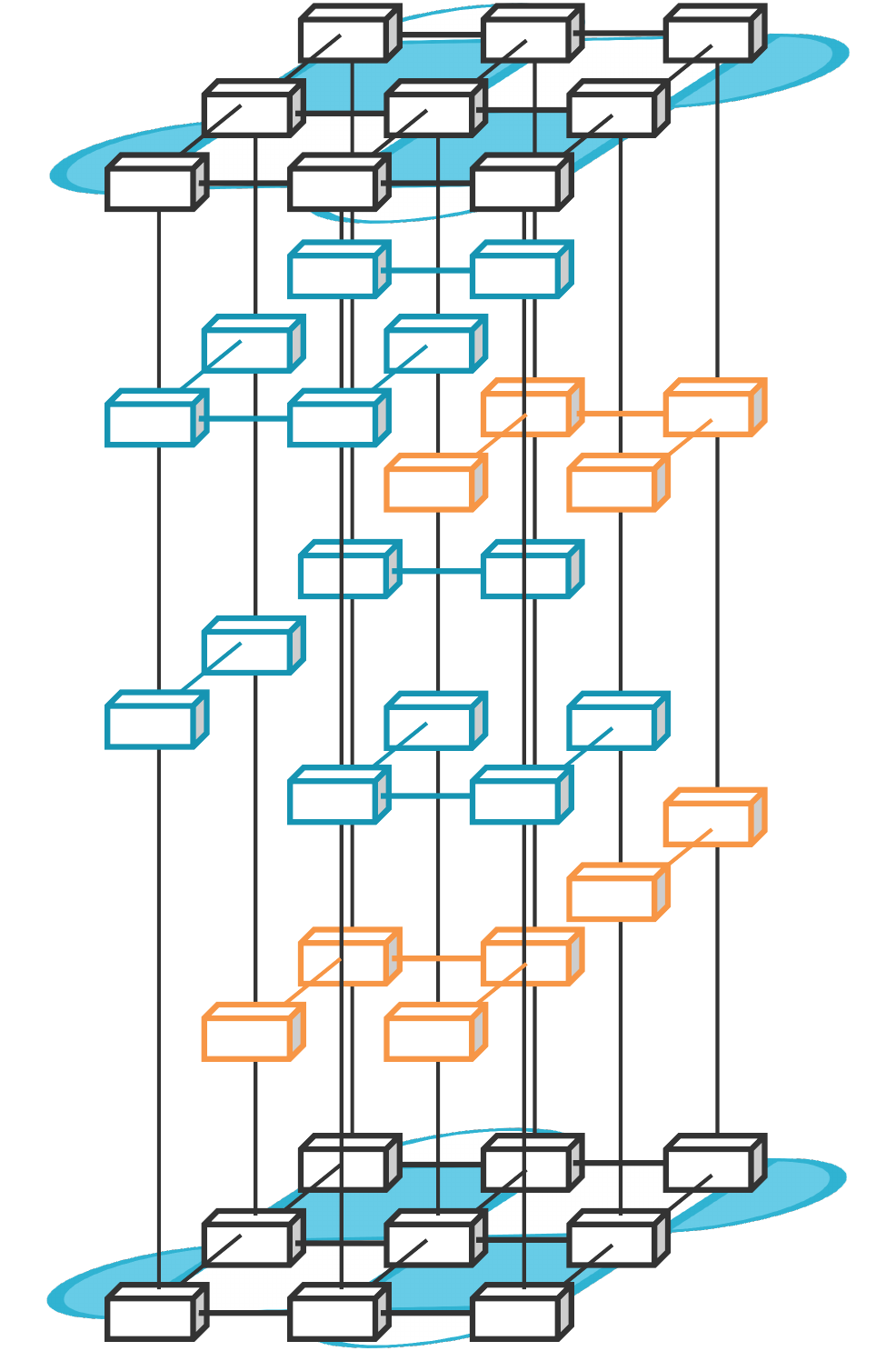}
        \end{overpic}
        \put(-120,190){{\textsf{(c)}}}
    \end{minipage}
    \caption{(a) TN representation of projectors $\Pi_{\pm g}=(I\pm g)/2$.
        (b) TN representation of $d=5$ initial code state $\ket{0}_L$. It can be made from $\ket{0}^{\otimes 25}$ by operating the $X$ stabilizer projectors. It is because $\ket{0}^{\otimes 25}$ is already stabilized by the products of $Z$.
        (c) A conceptual picture of the TN diagram of the likelihood $p(\bm{m}|\bm{\alpha})$ calculation of $d=3$ surface code.}\label{fig:TN}
\end{figure}

To efficiently simulate the surface code using TNs, it is essential to construct a TN representation of states or their processes.
In QEC procedures, the projectors onto the eigenspace of the stabilizer generators are crucial because they are used to obtain code states and to simulate syndrome measurements.
For a given stabilizer generator $g$, this projector is given by $\Pi_{\pm g} =(I \pm g)/2$, where $I$ is the identity operator. Each projector can be encoded as a tensor with bond dimension two, as illustrated in \reffig{fig:TN}(a).
When we label the horizontal leg indices by $\bm{k}$ connected by other tensors and the vertical leg indices by $i$ and $j$ for the physical qubit indices, element of the each tensor $M_{ij\bm{k}}$ is written as $M_{ij\bm{k}}=\delta_{\bm{k},\bm{0}}I_{ij}+\delta_{\bm{k},\bm{1}}\sigma_{ij}$, where $\sigma_{ij}$ is the Pauli operator acting on the physical qubit $i$ and $j$ corresponding to the stabilizer generator $g$.
Thus, the TN represent the product of $I$ for $\bm{k}=\bm{0}$, the product of $X$ or $Z$ for $\bm{k}=\bm{1}$ following the stabilizer generator $g$, and zero otherwise.
All legs have bond dimension two, so the representation is very efficient in terms of tensor network representation.

In the code-capacity noise model simulation, the whole process of surface code simulation can be written as $\mathcal{D}\circ\mathcal{M}\circ\mathcal{E}(\rho_{0})$, where $\rho_{0}$ is the initial code state such as $\ket{0}_{L}\bra{0}$, $\mathcal{E}$ is the noise process, $\mathcal{M}$ is the process of syndrome measurements, and $\mathcal{D}$ is the Pauli correction of decoding process.
Using the projectors $\Pi_{\pm g}$, we can represent the initial code states $\ket{0}_{L}$ (see \reffig{fig:TN}(b)).
The noise process $\mathcal{E}$ can be represented as the CPTP map of the density matrix and the process can be represented by the tensor.
The decoding process $\mathcal{D}$ is the Pauli operation and can also be represented by local tensors without any horizontal legs.
Within the code-capacity noise model, where measurement qubits are assumed to be noiseless, the syndrome measurement process $\mathcal{M}$ can be simulated by sequentially applying the projector tensors $\Pi_{\pm g}$, without the need to explicitly represent the measurement qubits in the TN.
Each syndrome measurement is simulated by evaluating $p_{+g}=Tr(\Pi_{+g} \rho \Pi_{+g})=Tr(\Pi_{+g} \rho)$, which gives the probability of the +1 outcome for the stabilizer generator $g$ in the current quantum state $\rho$.
The trace operation $Tr(\cdot)$ corresponds to the contraction of the whole TNs.
A random number $r$ uniformly distributed in $[0,1]$ is generated to simulate the measurement outcome: if $r < p_{+g}$, the outcome is assigned as $+1$; otherwise, it is $-1$.
The post-measurement state is then updated as $\rho \leftarrow \Pi_{+g}\rho/p_{+g}$ or $\rho \leftarrow \Pi_{-g}\rho/(1-p_{+g})$, according to the outcome.
By repeating the calculation for all $g\in\mathcal{S}_{g}$, the whole syndrome measurement process are completed, and we obtain the syndrome measurement outcomes $\bm{m}$ where $m_i$ is the syndrome measurement outcome of the $i$th stabilizer generator $g_{i}$.
By omitting the explicit representation of measurement qubits, the syndrome measurement process can be simulated efficiently within the tensor network formalism.

Applying this calculation method, we can obtain the likelihood function $p(\bm{m}|\bm{\alpha})$ where $\bm{\alpha}$ are the parameters representing the noise models $\mathcal{E}_{\bm{\alpha}}$.
The likelihood function is expressed as $p(\bm{m}|\bm{\alpha})=Tr({\displaystyle\prod_{i}}\Pi_{(m_{i} g_{i})}\mathcal{E}_{\bm{\alpha}}(\rho_{0}))$ so it can be calculated by preparing the TN representing the initial code state, contracting the tensor networks representing noise process $\mathcal{E}_{\bm{\alpha}}$ and the whole projector tensors corresponding to the measurement outcomes $\bm{m}$, and contracting whole TN like \reffig{fig:TN}(c).

The time complexity for the exact calculation of the entire TN by PEPS is $O(d^2 4^{d})$.
For large-scale surface code, we have to introduce some approximation.
By first contracting physical legs, the TN forms a 2D square network, allowing us to use the boundary MPS-MPO calculation techniques: the 2D square TNs can be regarded as multiple layers of MPO sandwiched between MPS boundary layers.
In our simulations, we adopt a low-rank approximation by truncating the bond dimension to a maximum value $\chi$ during each contraction of an MPS with an MPO layer. 
This approximation reduces the computational cost to $O(d^2 \chi^3)$, where $\chi$ is the maximum bond dimension. 
This allows surface code simulations to scale nearly linearly with the number of qubits.
This simulator has been applied for the decoding~\cite{Darmawan2018,darmawan2024optimal}, and we used it to evaluate improvements of decoding performance with our noise model estimator.
Furthermore, this simulator is capable of handling not only Pauli noise but also arbitrary noise, provided it can be expressed using TN, making it well-suited for our problem settings of general noise model estimation.

In the present formulation, several points should be noted.
The simulations explained here are based on the code-capacity noise model, which assumes perfect, noise-free syndrome measurements.
However, in real devices, noise also affects the syndrome measurements of ancilla qubits.
Thus, the current simulations are simplified by ignoring noise in syndrome measurements.
Incorporating more realistic noise models, such as phenomenological or circuit-level noise models, requires modifications to the TN simulations, making them more complex.
Specifically, we need to incorporate both tensors representing the measurement qubits and their noise processes. 
For noise model estimation, we must contract tensors corresponding to multi-cycle syndrome measurements to calculate the likelihood function by adding the tensors in \reffig{fig:TN}, which is similar to the corresponding modification for error correction process.
While attention must be paid to the practical implementation of the TN simulator, these modifications do not change its scaling with the number of qubits and may not significantly degrade its practical utility.

%% file: qne.tex
\section{Methods}\label{sec:qne}
\subsection{Bayesian inference of noise model based on syndrome measurements statistics}
First, we have to set objective functions to perform estimations.
In our case, the results of syndrome measurements, $\bm{m}_{0:n-1}=\{\bm{m}_{0}, \bm{m}_{1}, \cdots, \bm{m}_{n-1}\}$, are only the information we have.
Here, $n$ is the number of cycles of syndrome measurements, and $\bm{m}_{i}$ is the set of one cycle syndrome measurement results, so if we consider $L\times W$ surface code, the number of elements in $\bm{m}_{i}$ is $L\times W-1$.
The noise model is parameterized by a set of parameters, $\bm{\alpha}$, even in the case of general noise models.
From the viewpoint of Bayesian inference, estimation is performed using the conditional probability distribution $p(\bm{\alpha}|\bm{m}_{0:n-1})$.
The conditional probability $p(\bm{\alpha}|\bm{m}_{0:n-1})$ is also called the posterior distribution in Bayesian inference.
We calculate the posterior distribution using the Bayes' theorem:
\begin{align}
    \begin{split}
        p(\bm{\alpha}|\bm{m}_{0:n-1}) & = \frac{p(\bm{m}_{0:n-1}|\bm{\alpha})p(\bm{\alpha})}{p(\bm{m}_{0:n-1})}                                                            \\
                                      & = \frac{p(\bm{m}_{0:n-1}|\bm{\alpha})p(\bm{\alpha})}{\displaystyle \int d\bm{\alpha}\,p(\bm{m}_{0:n-1}|\bm{\alpha})p(\bm{\alpha})} \\
                                      & \propto p(\bm{m}_{0:n-1}|\bm{\alpha})p(\bm{\alpha}),
    \end{split}
    \label{eqs: Bayesian theorem}
\end{align}
where $p(\bm{\alpha})$ is the prior distribution and $p(\bm{m}_{0:n-1}|\bm{\alpha})$ is the likelihood function.
In the Bayesian inference, we initially set the prior distribution $p(\bm{\alpha})$, and by multiplying this with the likelihood $p(\bm{m}_{0:n-1}|\bm{\alpha})$, we refine the posterior function and enhance the estimation results.
The prior distribution can be chosen based on prior knowledge of the noise model. 
In this work, we employ a uniform prior over a restricted range to avoid unpractical solutions.
Certain parameters yield identical syndrome statistics and they are indistinguishable from a statistical perspective.
For example, noise models with probability $p$ of $X$ Pauli errors yield the same syndrome statistics as ones with probability $1-p$. 
Thus, without an appropriate prior, both solutions are indistinguishable from a statistical perspective. 
By the prior, we can select the solution that is more practically reasonable, in this case, the one with $p\leq 0.5$.

After calculating the posterior distribution, we can analyze the estimation results by observing the whole landscape of the distribution or statistics computed from the distribution.
We have several choices for statistics, but we chose the expected a posterior~(EAP) estimator for evaluating results, which is defined as $\bm{\alpha}^{*}=\int d\bm{\alpha} \, \bm{\alpha} p(\bm{\alpha}|\bm{m}_{0:n-1})$.

\subsection{Markov chain Monte Carlo for stationary noise model}\label{subsec:MCMC}
In practical situations, the posterior distribution $p(\bm{\alpha}|\bm{m})$ cannot be evaluated analytically.
Therefore, sampling-based methods must be employed for its estimation. 
Among these, MCMC sampling is a fundamental and particularly effective tool for Bayesian inference. 
The most widely used MCMC algorithm is the Metropolis–Hastings (M-H) algorithm~\cite{Nicholas1953}, which allows sampling from an unnormalized target distribution. 
In the context of Bayesian inference, we can sample from the posterior distribution without knowing the normalization factor, $p(\bm{m}_{0:n-1})$ but only the likelihood $p(\bm{m}_{0:n-1}|\bm{\alpha})$ and the prior distribution $p(\bm{\alpha})$ according to Eq.~\ref{eqs: Bayesian theorem}.
In our methods, we calculate the likelihood using the TN simulation.
Generally, the time complexity of approximate TN calculation is $O(\# qubits)$, so the Bayesian inferences based on MCMC sampling are efficient.
In the present paper, we used the simulator for the surface code in \refsec{subsec:TN simulator}.
The whole algorithm is given in \refalg{alg:qee MCMC}.
Here, the burn-in time $s$ refers to the number of initial samples that are discarded to ensure that the remaining samples are representative of the posterior distribution. The proposal distribution $q(\bm{\alpha}',\bm{\alpha})$ is used to generate candidate states in the M–H algorithm.
\begin{algorithm}[H]
    \caption{Bayesian inference using MCMC for stationary noise model}\label{alg:qee MCMC}
    \begin{algorithmic}[1]
        \Require{results of syndrome $\bm{m}_{0:n-1}$, burn-in-time $s$,\\ total MC steps $T$ $(>s)$, prior distribution $p(\bm{\alpha})$, \\proposal distribution $q(\bm{\alpha}',\bm{\alpha})$, \\max bond dimension of TN $\chi$}
        \Ensure{samples of $p(\bm{\alpha}|\bm{m}_{0:n-1})$}
        \State Initialize $\bm{\alpha}^{0}\sim p(\bm{\alpha})$
        \State Likelihood $p(\bm{m}_{0:n-1}|\bm{\alpha}^{0})\gets TNsimulator(\bm{m}_{0:n-1},\bm{\alpha}^{0},\chi)$
        \For{$i=0,1,\ldots$, T}
        \State Sample $\bm{\alpha}'\sim q(\bm{\alpha}',\bm{\alpha}^{i})$
        \State Likelihood $p(\bm{m}_{0:n-1}|\bm{\alpha}')\gets TNsimulator(\bm{m}_{0:n-1},\bm{\alpha}',\chi)$
        \State $\gamma\gets \min\left(1,\frac{p(\bm{m}_{0:n-1}|\bm{\alpha}')}{p(\bm{m}_{0:n-1}|\bm{\alpha}^{i})}\right)$
        \State Sample $r\sim \text{Uniform}(r;0,1)$
        \If{$r<\gamma$}
        \State $\bm{\alpha}^{i+1}\gets \bm{\alpha}'$
        \State $p(\bm{m}_{0:n-1}|\bm{\alpha}^{i+1}) \gets p(\bm{m}_{0:n-1}|\bm{\alpha}')$
        \Else
        \State $\bm{\alpha}^{i+1}\gets \bm{\alpha}^{i}$
        \State $p(\bm{m}_{0:n-1}|\bm{\alpha}^{i+1}) \gets p(\bm{m}_{0:n-1}|\bm{\alpha}^{i})$
        \EndIf
        \EndFor
        \State \Return $\{\bm{\alpha}^{s},\bm{\alpha}^{s+1},\ldots,\bm{\alpha}^{T}\}$
    \end{algorithmic}
\end{algorithm}
In this algorithm, we set the proposal distribution as $q(\bm{\alpha'},\bm{\alpha})=q(\bm{\alpha},\bm{\alpha'})$ so that the detailed balance condition is satisfied.
The detailed balance condition requires that the transition probabilities between any two states are balanced, which guarantees convergence to the target distribution.
This method can be applied to more general noise models than Pauli noise models, as the likelihood can be calculated by the TN simulator.

There are several important considerations for our MCMC and TN methods. 
One is that, unlike perfect sampling, MCMC samples generally exhibit autocorrelation between samples obtained in the same chain.
So, we have to collect many samples and thin out samples to reduce the dependency of initial distribution.
Second is the error in the estimates.
For MCMC, there are some statistical errors.
Furthermore, for TN, there is a truncation error in the approximate contraction of the TN.
To evaluate the accuracy of the estimates, we have to consider both of them.
Third, the MCMC method is not suitable for the time-varying noise models which are more realistic in quantum devices~\cite{EtxezarretaMartinez2021,Kilmov2018fluctuations}.
To deal with them, we developed another method based on the SMC.

\begin{figure}[tbp]
    \centering
    \includegraphics[width=1.\hsize]{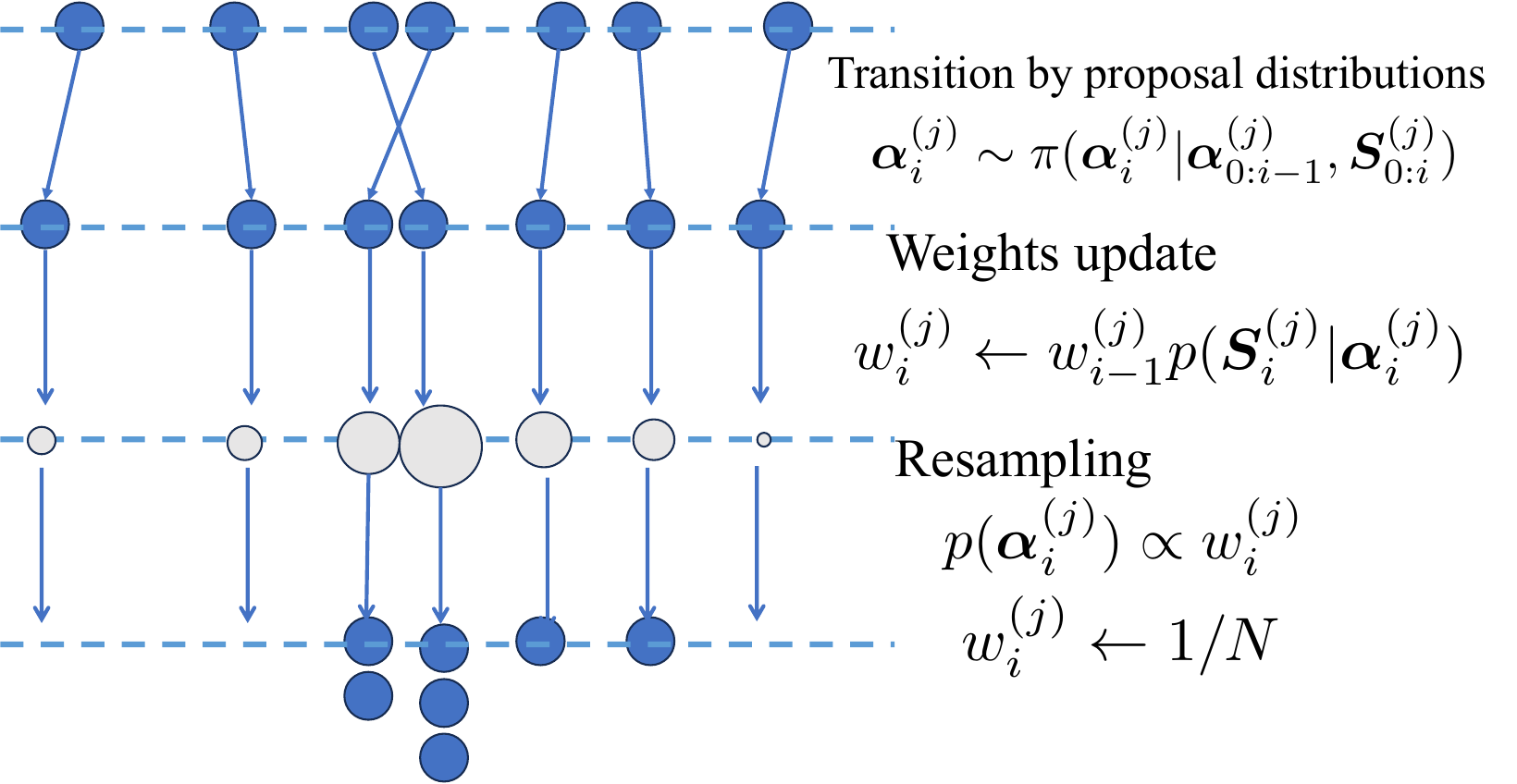}
    \caption{
        Conceptual diagram of updates of weighted particles for one cycle results in SMC.
    }\label{fig:SMC}
\end{figure}

\subsection{Sequential Monte Carlo for time-varying noise model\label{subsec:SMC}}
In QEC, we can get results of syndrome measurements sequentially.
The form of QEC and the decode performance are influenced by the information of the noise model.
However, in realistic situations, the noise models vary over time.
So, continuously updating the noise model information is essential to improve QEC performance.
Such sequential estimations can be realized by the SMC~\cite{Doucet2001, Kitagawa1993, Gordon1993}, which is also called the particle filter.

First, we explain the problem setting for time-varying noise model estimation.
For the time-varying case, the noise model parameters are described as $\{\bm{\alpha}_{0},\bm{\alpha}_{1},\ldots,\bm{\alpha}_{n-1}\}$ and we assume the Markov process, that is, each parameter depends only on its immediate predecessor.
Syndrome measurements $\bm{m}_{i}$ can be obtained sequentially which are generated according to $p(\bm{m}_{i}|\bm{\alpha}_{i})$.
Our goal is to estimate the latest $\bm{\alpha}_{i}$ based on the sequence of previous syndrome measurement results $\bm{m}_{0:i}=\{\bm{m}_{0},\bm{m}_{1},\ldots,\bm{m}_{i}\}$.
In the Bayesian inference, this is achieved by calculating the posterior distribution $p(\bm{\alpha}_{0:i}|\bm{m}_{0:i})$ and the marginal distribution $p(\bm{\alpha}_{i}|\bm{m}_{0:i})$.

To simulate the posterior distribution continuously, we use the SMC.
In the SMC, the weighted particles are utilized to represent the posterior distribution $p(\bm{\alpha}_{0:i}|\bm{m}_{0:i})$ as a product of weight function $w(\bm{\alpha}_{0:i})$, called importance weight, and proposal distribution $\pi(\bm{\alpha}_{0:i}|\bm{m}_{0:i})$ like
\begin{equation}
    p(\bm{\alpha}_{0:i}|\bm{m}_{0:i})= w(\bm{\alpha}_{0:i})\pi(\bm{\alpha}_{0:i}|\bm{m}_{0:i}).
\end{equation}
Thus, $N$ weighted particles $\{\bm{\alpha}_{0:i}^{(j)}\}$ ($j=1,\ldots,N$) sampled from $\pi(\bm{\alpha}_{0:i}|\bm{m}_{0:i})$ with weights $w(\bm{\alpha}_{0:i})$ can be used to approximate the posterior distribution $p(\bm{\alpha}_{0:i}|\bm{m}_{0:i})$.
Here, the proposal distribution $\pi(\bm{\alpha}_{0:i}|\bm{m}_{0:i})$ can be chosen according to prior knowledge or convenience, while the weight function $w(\bm{\alpha}_{0:i})$ is updated based on the observed data.
When evolving the particles according to the proposal distribution with obtained syndrome measurement results recursively like
\begin{align}
    \begin{split}
        \pi(\bm{\alpha}_{0:i}|\bm{m}_{0:i}) & = \pi(\bm{\alpha}_{0:i-1}|\bm{m}_{0:i-1})\pi(\bm{\alpha}_{i}|\bm{\alpha}_{0:i-1},\bm{m}_{0:i}) \\
                                            & = \pi(\bm{\alpha}_0)\prod_{j=0}^{i-1}\pi(\bm{\alpha}_{j+1}|\bm{\alpha}_{0:j},\bm{m}_{0:j+1}),
    \end{split}
\end{align}
the importance weight $w(\bm{\alpha}_{0:i})$ will be updated recursively as
\begin{align}
    \begin{split}
        w(\bm{\alpha}_{0:i}) & = \frac{p(\bm{\alpha}_{0:i}|\bm{m}_{0:i})}{\pi(\bm{\alpha}_{0:i}|\bm{m}_{0:i})}    \\
                             & = w(\bm{\alpha}_{0:i-1})\frac{p(\bm{\alpha}_{i},\bm{m}_{i}|\bm{\alpha}_{0:i-1},\bm{m}_{0:i-1})}{p(\bm{m}_i|\bm{m}_{0:i-1})\pi(\bm{\alpha}_{i}|\bm{\alpha}_{0:i-1},\bm{m}_{0:i})} \\ 
                             & \propto w(\bm{\alpha}_{0:i-1})\frac{p(\bm{m}_{i}|\bm{\alpha}_{i})p(\bm{\alpha}_{i}|\bm{\alpha}_{i-1})}{\pi(\bm{\alpha}_{i}|\bm{\alpha}_{0:i-1},\bm{m}_{0:i})},
    \end{split}
\end{align}
which is derived from the Markov process of $\bm{m}_i$ and $\bm{\alpha}_i$. 
This allows for rapid calculation of weights using the likelihood.

To consider such a sequential weight update, we have to deal with excessive weight degeneracy.
In SMC, there is a well-known approach, resampling.
There are four famous resampling methods: the multinomial, stratified, systematic, and residual resampling~\cite{Gordon1993, Kitagawa1996, Carpenter1999, Beadle1997, Liu1998}.
Their time complexities are $O(N)$ or $O(N\log N)$.
Among various resampling methods, the recommendation for the first choice is systematic resampling because it exhibits good performance in many cases~\cite{Douc2005, Li2015}.
In the present work, we mainly use the systematic resampling method.
The systematic resampling algorithm is given in \refalg{alg:systematicresampling}.
\begin{algorithm}[H]
    \caption{Systematic resampling}
    \label{alg:systematicresampling}
    \begin{algorithmic}[1]
        \Require{weights $\{w^{(j)}\}$, samples $\{\bm{x}^{(j)}\}$}
        \Ensure{samples $\{\bm{x'}^{(j)}\}$}
        \State $\{C^{j}\}\gets$ Cumulative sum of $\{w^{(j)}\}$
        \State Number of samples $n$, Systematic choice $u_{0}\sim \text{Uniform}(u;0,1/n)$
        \State Index $i\gets 0,j\gets 0$
        \While{$i<n$}
        \While{$C^{j}<u_{0}+i/n$}
        \State $j\gets j+1$
        \EndWhile
        \State $\bm{x'}^{i}\gets \bm{x}^{j}$
        \State $i\gets i+1$
        \EndWhile
        \State \Return $\{\bm{x'}^{(j)}\}$
    \end{algorithmic}
\end{algorithm}

Its conceptual diagram is shown in \reffig{fig:SMC} to facilitate an understanding of one cycle evolution of weighted particles.
If the proposal distribution $\pi(\bm{\alpha}_{i}|\bm{\alpha}_{0:i-1},\bm{m}_{0:i})$ equals to the prior distribution $p(\bm{\alpha}_{i}|\bm{\alpha}_{i-1})$, the weight will be $w(\bm{\alpha}_{0:i})\propto w(\bm{\alpha}_{0:i-1})p(\bm{m}_{i}|\bm{\alpha}_{i})$ and we can calculate the weight by only using the likelihood.
Such a choice of proposal distribution is called a bootstrap filter~\cite{gordon1993novel,doucet2000sequential} and is often used in realistic situations.
It is used in our method.

Our method for estimating time-varying noise is based on SMC and the bootstrap filter.
To use the bootstrap filter, we must choose the proposal distribution according to the system models of applied noise.
Due to experimental efforts, the noise model parameters are regulated to ensure gradual changes rather than abrupt fluctuations.
Using a random walk distribution as the proposal distribution is a common approach for tracking slowly varying parameters and is well-suited to our scenario.
\begin{algorithm}[H]
    \caption{Bayesian inference using SMC for time-varying noise model}\label{alg:qee SMC}
    \begin{algorithmic}[1]
        \Require{results of syndrome $\bm{m}_{0:n-1}$, resampling interval $T$,\\number of particles $N$,\\ smoothing time $t_s$, proposal distribution $q(\bm{\alpha}',\bm{\alpha})$,\\ prior distribution $p(\bm{\alpha})$, max bond dimension of TN $\chi$}
        \Ensure{estimated parameters $\{\bar{\bm{\alpha}_{i}}\}$}
        \State Initialize $N$ particles from prior distribution $\{\bm{\alpha}_{0}^{(j)}\}\sim p(\bm{\alpha})$
        \State Initialize weights $\{w_{0}^{(j)}\gets 1/N\}$
        \State EAP estimator ${\bm{\alpha}^{EAP}}_{0}\gets \text{WeightedMean}(\{\bm{\alpha}_{0}^{(j)}\},\{w_{0}^{(j)}\})$
        \For{$i=1,\ldots$}
        \For{$j=0,1,\ldots,N$}
        \State Sample $\bm{\alpha}_{i}^{(j)}\sim q(\bm{\alpha}_{i}^{(j)},\bm{\alpha}_{i-1}^{(j)})$
        \State $w_{i}^{(j)} \gets w_{i-1}^{(j)}\text{TNsimulator}(\bm{m}_{i},\bm{\alpha}^{(j)}_i,\chi)$
        \EndFor
        \If{$i\% T = 0$}
        \State Normalize weights $\{w_{i}^{(j)}\}\gets \text{Normarize}(\{w_{i}^{(j)}\})$
        \State resampling $\{\bm{\alpha}_{i}^{(j)}\}\gets \text{Resample}(\{\bm{\alpha}_{i}^{(j)}\},\{w_{i}^{(j)}\})$
        \State Update weights $\{w_{i}^{(j)}=1/N\}$
        \EndIf
        \State EAP estimator ${\bm{\alpha}^{EAP}}_{i}\gets \text{WeightedMean}(\{\bm{\alpha}_{i}^{(j)}\},\{w_{i}^{(j)}\})$
        \State Estimator $\bar{{\bm{\alpha}}_{i}}\gets \text{Mean}(\{{\bm{\alpha}^{EAP}}_{max(0,i-t+1)},\ldots,{\alpha^{EAP}}_{i}\})$
        \EndFor
        \State \Return $\{\bar{{\bm{\alpha}}_{i}}\}$
    \end{algorithmic}
\end{algorithm}

SMC has several advantages.
One of them is the ease of implementing parallel computing.
We can trivially parallelize the updating process of SMC by distributing particles to many processors.
On the other hand, the resampling process requires communication between processors.
Optimizing the interval for resampling based on some criterion unlike \reffig{fig:SMC} is recommended for efficiency.
An effective particle number $1/\sum_{j} (w_{i}^{(j)})^2$ is often used for the criterion of resampling, but it requires additional communication between processes.
In the present simulation, we use a constant resampling interval instead of the effective particle number.

On the other hand, SMC has some disadvantages.
One is the sampling bias inherent in the syndrome measurement results.
Additionally, the sampling bias inherent in syndrome measurements impacts the accuracy of our estimations.
The relation between syndrome measurements and noise models is not simple and probabilistic.
To reduce the sampling bias, we must accumulate many results from the syndrome measurement.
For that purpose, we introduce smoothing processes using near-past data: The estimate is evaluated by not only the EAP estimator of the latest particles but also the near-past EAP estimators.
To sum up, our method based on the bootstrap filter is described in \refalg{alg:qee SMC}.
Second is that SMC is based on the Markov process and, therefore, cannot be directly applied to non-Markovian noise models.
Unfortunately, there are some noise sources in quantum devices that are non-Markovian~\cite{white2020demonstration, white2022nonmarkovian,nonMarkovianDynamics2017}.
To accommodate such cases, our method must be extended to handle non-Markovian dynamics~\cite{Jacob2018bayesian,Naesseth2019elements}.
Although SMC is theoretically inapplicable to non-Markovian processes, in practice these effects can be approximated by augmenting the weighted particles with additional parameters that retain information from recent past configurations. 
The degree of non-Markovianity that can be captured depends on the depth of the incorporated temporal history.

\subsection{Comparison with previous works based on Pauli noise models}
In the previous works, there are some methods to estimate Pauli noise models~\cite{Wagner2021,Wagner2022,Hockings2025scalable}.
Of course, our methods can be applied to the Pauli noise models by the TN simulator to calculate the likelihood.
By restricting the noise model to Pauli noise, we can use the characterization of the stabilizer code and the Pauli noise models directly.
For example, in~\cite{Wagner2021}, the estimation is realized by combining EM algorithm and the belief propagation which uses the properties of the stabilizer code and Pauli noise models.
In~\cite{Wagner2022}, the estimability of the Pauli noise model based on the syndrome statistics is discussed precisely by Boolean Fourier analysis based on the calculus of Pauli noise channels.
In~\cite{Hockings2025scalable}, the estimation is realized by averaged circuit eigenvalue estimation which is applicable to the clifford circuits and Pauli noise models.
It can be said that such methods can offer faster estimation and more general estimability analysis than our framework based on MCMC and SMC by sacrificing the generality of the noise model.
The generalization of the noise model is one of the most significant advantages of our proposed methods and valuable because the noise model in quantum devices is not limited to Pauli noise~\cite{QCQI}.
However, it should be noted that the syndrome statistics depend on the initial logical information when considering the general noise models, which is not the case for the Pauli noise models.
In the case of Pauli noise models, the syndrome statistics are independent of the initial logical information, so the estimation can be performed without considering the initial logical information.
On the other hand, noise models estimations for general noise models must consider the initial logical information, which leads to more complex estimation processes or limits the practical usability.

There is another point to be noted about the comparisons between Pauli noise models and general noise models.
The point is the randomized compilation~\cite{Wallman2016}, and by using it, the general noise models are projected to the Pauli noise models, which is often pointed out as the reason why the Pauli noise estimations are enough.
Especially, for stabilizer codes, randomized compilations can be applied by just adding operations which do not require quantum gates operation on real devices but only classical processing of modifying the Pauli frame~\cite{riesebos2017pauli,Knill2005,Wallman2016}.
However, it needs additional discussion about its practical usefulness.
In principle, the exact randomized compilation requires exponential overhead in the number of noise channels.
In the works related to the randomized compilation~\cite{Wallman2016,Cai2019,Hashim2021randomized,Perrin2024mitigating}, they show successful results of the randomized compilation, but their experiments are limited to much smaller program size than one of fault-tolerant quantum computing.
It is not clear whether the randomized compilation is practically useful, and the general noise model estimation is still important to be developed.

%% file: result.tex
\begin{figure*}[tbp]
    \begin{minipage}[b]{0.32\linewidth}
        \centering
        \includegraphics[width=1.\hsize]{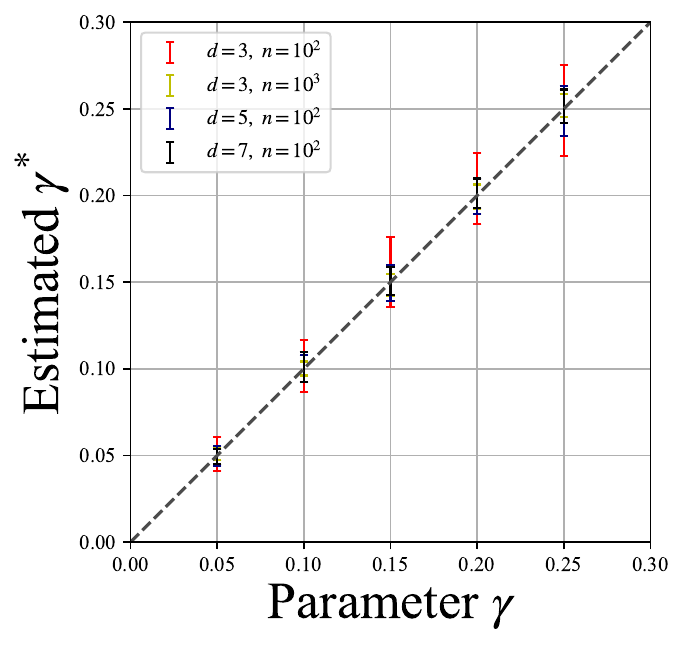}
    \end{minipage}
    \put(-150,155){{\textsf{(a)}}}
    \begin{minipage}[b]{0.32\linewidth}
        \centering
        \includegraphics[width=1.\hsize]{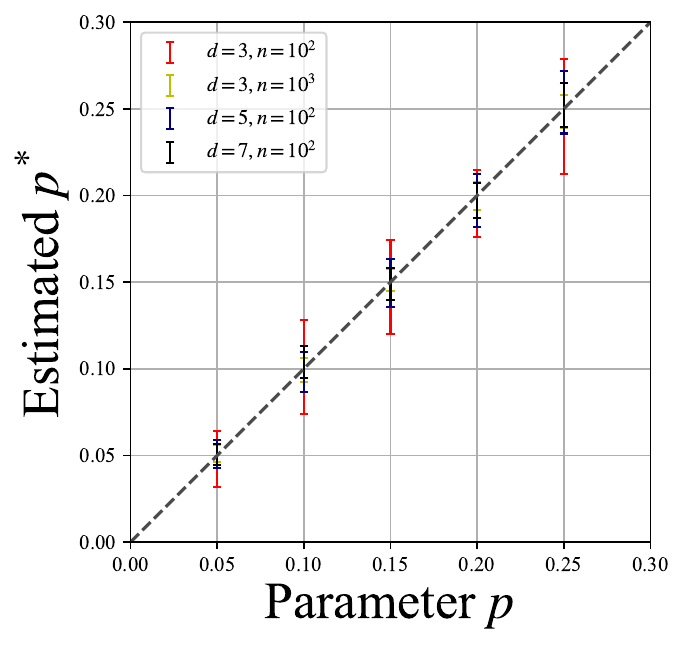}
    \end{minipage}
    \put(-150,155){{\textsf{(b)}}}
    \begin{minipage}[b]{0.32\linewidth}
        \centering
        \includegraphics[width=1.\hsize]{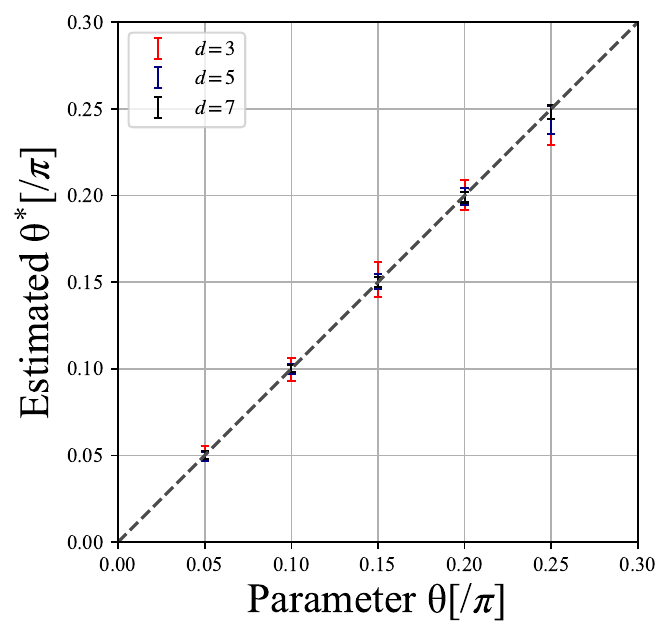}
    \end{minipage}
    \put(-150,155){{\textsf{(c)}}}
    \caption{
        Results of estimation for the one-parameter uniform noise models: (a) amplitude damping, (b) phase damping, and (c) systematic rotation.
    }\label{fig:one_para}
\end{figure*}

\section{Numerical experiments}\label{sec:result}
In this section, we show the results of numerical experiments with our proposed method, the Bayesian noise model estimation based on the TN simulation of the surface code, and the Monte Carlo sampling technique.

\subsection{Results of Markov chain Monte Carlo for stationary noise models}
We explain and discuss the results for the stationary noise models using MCMC explained in \refalg{alg:qee MCMC}.
This method is supposed to be used for the calibrations or discussion of estimability.
\subsubsection{One parameter uniform noise models}
First, we show the results for the uniform one-parameter noise model case.
We consider the three noise models: amplitude damping (AD), phase damping (dephase), and systematic rotation (SR) along the $z$-axis.
These three noise models are described as
\begin{align}
    \begin{split}
        \mathcal{E}_{AD}(\rho) & =K_{0}\rho K_{0}^{\dagger}+K_{1}\rho K_{1}^{\dagger}, \\
        K_{0}                  & =\ket{0}\bra{0} +\sqrt{1-\gamma}\ket{1}\bra{1},       \\
        K_{1}                  & =\sqrt{\gamma}\ket{0}\bra{1},
    \end{split}
\end{align}
\begin{gather}
    \begin{split}
        \mathcal{E}_{dephase}(\rho) & =K_{0}\rho K_{0}^{\dagger}+K_{1}\rho K_{1}^{\dagger}, \\
        K_{0}                       & =\ket{0}\bra{0}+\sqrt{1-p}\ket{1}\bra{1},             \\
        K_{1}                       & =\sqrt{p}\ket{1}\bra{1},
    \end{split}
\end{gather}
and
\begin{align}
    \mathcal{E}_{SR}(\rho)=e^{-i\theta Z}\rho e^{i\theta Z},
\end{align}
respectively.
The size of the surface code is $d\times d$, which has the code distance $d$, and $n$ denotes the number of sets of syndrome measurements.
In these numerical experiments, the initial state is set to $\rho_{0}\propto \ket{0}_{L}\bra{0} + \ket{1}_{L}\bra{1}$.
The estimation results are summarized in \reffig{fig:one_para}.

\begin{figure}[tbp]
    \centering
    \includegraphics[width=0.7\hsize]{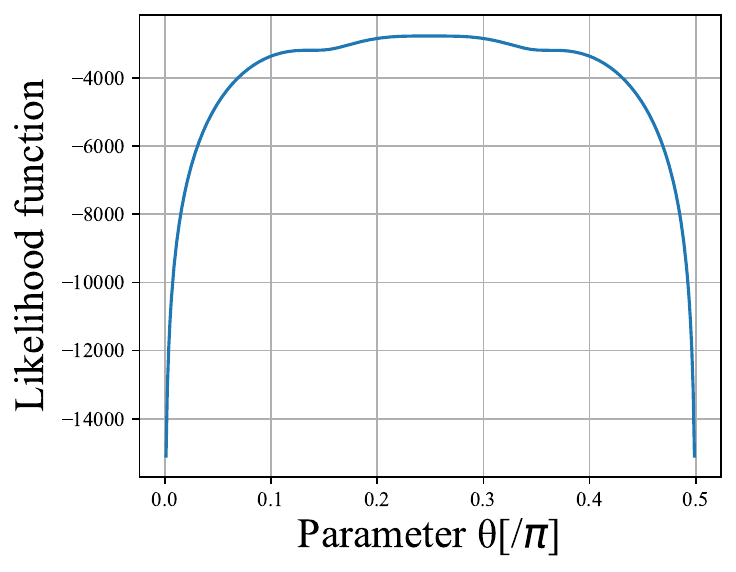}
    \caption{
        Likelihood function $p(\bm{m}|\theta)$ of the systematic rotation noise model calculated for the samples $\bm{m}\sim p(\bm{m}|\theta=0.25\pi)$.
    }\label{fig:SRlikelihood}
\end{figure}

We set the burn-in time and the total MCMC steps to $500$ and $10^3$, respectively.
Independent estimations for each parameter were performed 20 times for $d=7$ and 50 times for the other cases to estimate the sampling bias of syndrome measurements.
The TN simulator did not use any approximation, so this is the optimal estimation for TN contraction approximation.
We find that all the estimations are successful, as seen in \reffig{fig:one_para}.
The error bars in the figures mainly represent the sampling bias of syndrome measurements.
The size of the error bars should be inversely proportional to the $\sqrt{n}$, where $n$ is the total count of syndrome measurements.
We can confirm this behavior in the figures.

The estimation results for the systematic rotation have smaller errors compared with the others.
It comes from the difference in noise parameters representing noise models.
The range of noise model parameters $\beta$ of amplitude and phase damping models are $0\leq\beta\leq 1$.
On the other hand, the noise model parameter of systematic rotation has cycle $\pi /2$.
This is why systematic rotation error bars are smaller than the other noise models.
Second, for some initial values, the estimation sometimes fails.
The reason is that the likelihood function for the systematic rotation noise models is multimodal, as shown in \reffig{fig:SRlikelihood}.
The estimation based on the MCMC and EAP estimator is not suitable for multimodal functions.
In such situations, we must set the prior distribution to let the posterior distribution be a single modal.

\begin{figure}[tbp]
    \centering
    \includegraphics[width=0.7\hsize]{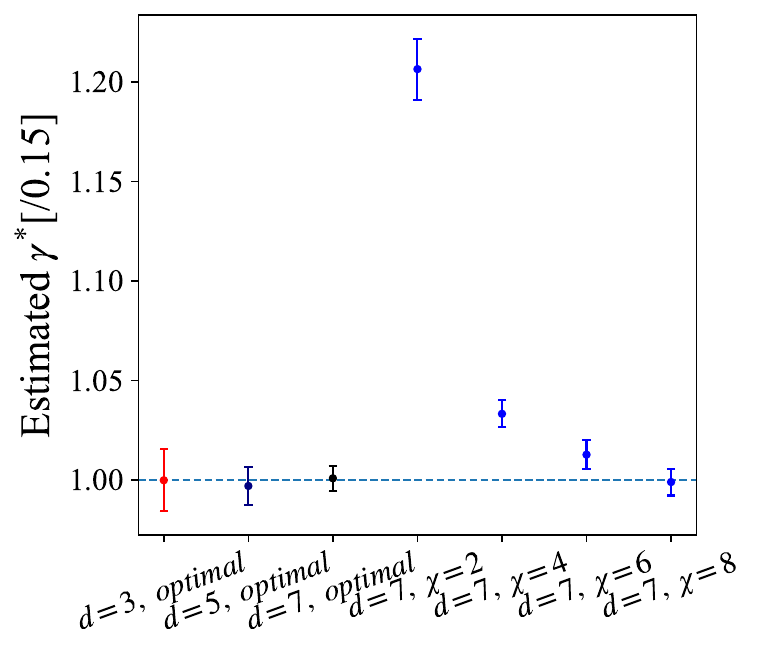}
    \caption{
        Dependence of estimation results on code size and maximum bond dimension for amplitude damping with $\gamma = 0.15$.
        All plotted values are divided by $0.15$.
    }\label{fig:AD015}
\end{figure}

To discuss the accuracy of the tensor network approximation, we did extra experiments for amplitude damping with noise model parameter 0.15.
The results are presented in \reffig{fig:AD015}.
The number of sets of syndrome measurements is $n=10^2$, and $\chi$ is the maximum bond dimension for the low-rank approximation of TN contraction.
We can see clearly that the larger code has smaller errors.
As for the effect of maximum bond dimension, we can see a systematic bias for small $\chi$.
However, as we increase $\chi$, the low-rank approximation becomes more and more accurate, and the bias is negligible for $\chi=8$ in this case.
The required bond dimension shown in \reffig{fig:AD015} is consistent with the previous works~\cite{Darmawan2017,Darmawan2018}, which demonstrate that even for larger system sizes, such as $d=9$ and $d=11$, the bond dimension remains stable at around $\chi=8$.
This suggests that the bond dimension does not significantly increase with larger system sizes, indicating that the TN contraction can be efficiently performed even for larger codes.
Therefore, while careful consideration of the required bond dimension for noise model estimation is necessary, the required bond dimension is expected to be manageable for practical applications.

\subsubsection{Two parameter uniform noise models}
In this section, we try to estimate the two-parameter noise model cases.
We consider AD+dephase noise models and generalized amplitude damping (GAD) noise models.
AD$+$dephase is the combination of the amplitude damping and phase damping noise models, and the CPTP map is described as $\mathcal{E}_{dephase}\circ\mathcal{E}_{AD}(\rho)$.
GAD is defined as
\begin{align}
    \begin{split}
        \mathcal{E}_{GAD}(\rho) & =\sum_{i=0}^{3}K_{i}\rho K_{i}^{\dagger},                  \\
        K_{0}                   & =\sqrt{p}(\ket{0}\bra{0}+\sqrt{1-\gamma}\ket{1}\bra{1}),   \\
        K_{1}                   & =\sqrt{1-p}(\sqrt{1-\gamma}\ket{0}\bra{0}+\ket{1}\bra{1}), \\
        K_{2}                   & =\sqrt{\gamma p}\ket{0}\bra{1},                            \\
        K_{3}                   & =\sqrt{\gamma(1-p)}\ket{1}\bra{0}.
    \end{split}
\end{align}
\begin{figure}[tbp]
    \centering
    \includegraphics[width=0.7\hsize]{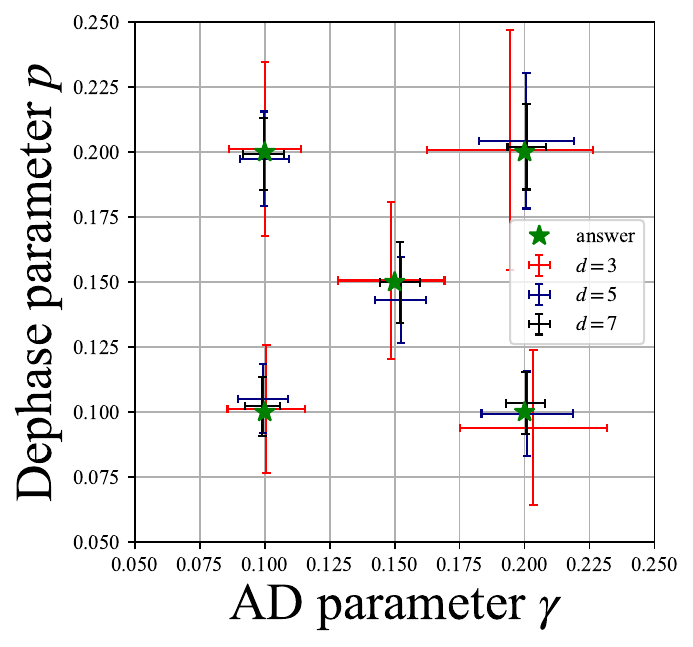}
    \caption{
        Results of estimation of the AD$+$dephase noise model.
        The green stars in the figures represent the true model parameters.
    }\label{fig:ADdephase}
\end{figure}
\begin{figure}[tbp]
    \begin{minipage}[b]{0.45\linewidth}
        \centering
        \begin{overpic}[keepaspectratio,width=\hsize]{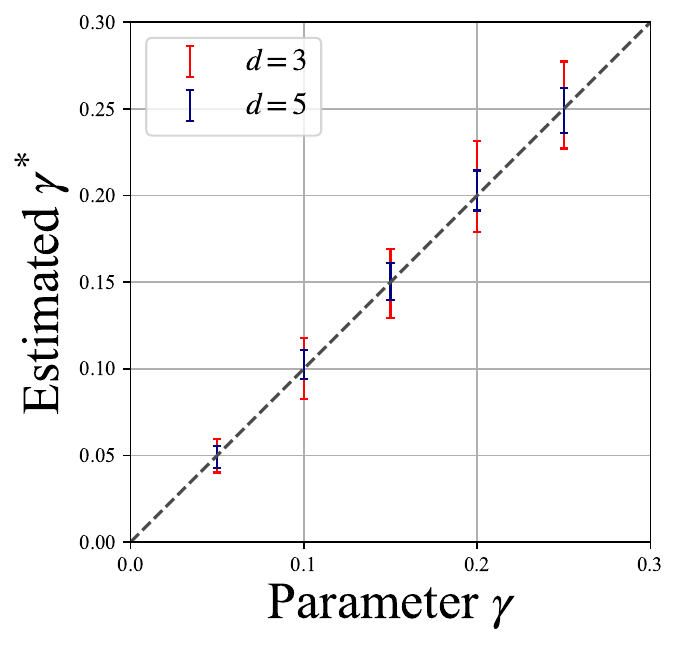}
        \end{overpic}
        \put(-105,105){{\textsf{(a)}}}
    \end{minipage}
    \begin{minipage}[b]{0.45\linewidth}
        \centering
        \begin{overpic}[keepaspectratio,width=\hsize]{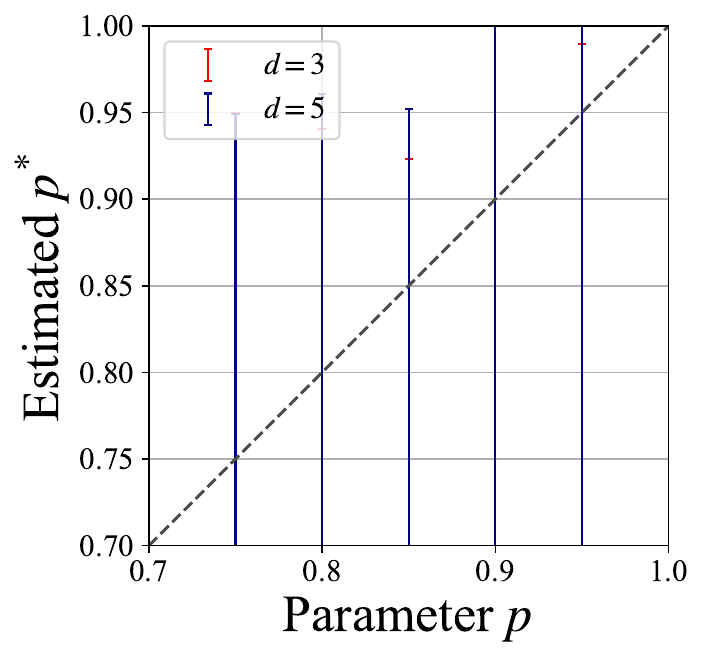}
        \end{overpic}
        \put(-105,105){{\textsf{(b)}}}
    \end{minipage}
    \caption{
        Results of one parameter estimation for the GAD noise model parameters:
        (a) Estimation of $\gamma$ with fixed $p=0.9$. (b) Estimation of $p$ with fixed $\gamma=0.1$.
    }\label{fig:GAD}
\end{figure}

\begin{figure}[tbp]
    \begin{minipage}[b]{0.48\linewidth}
        \centering
        \includegraphics[width=\hsize]{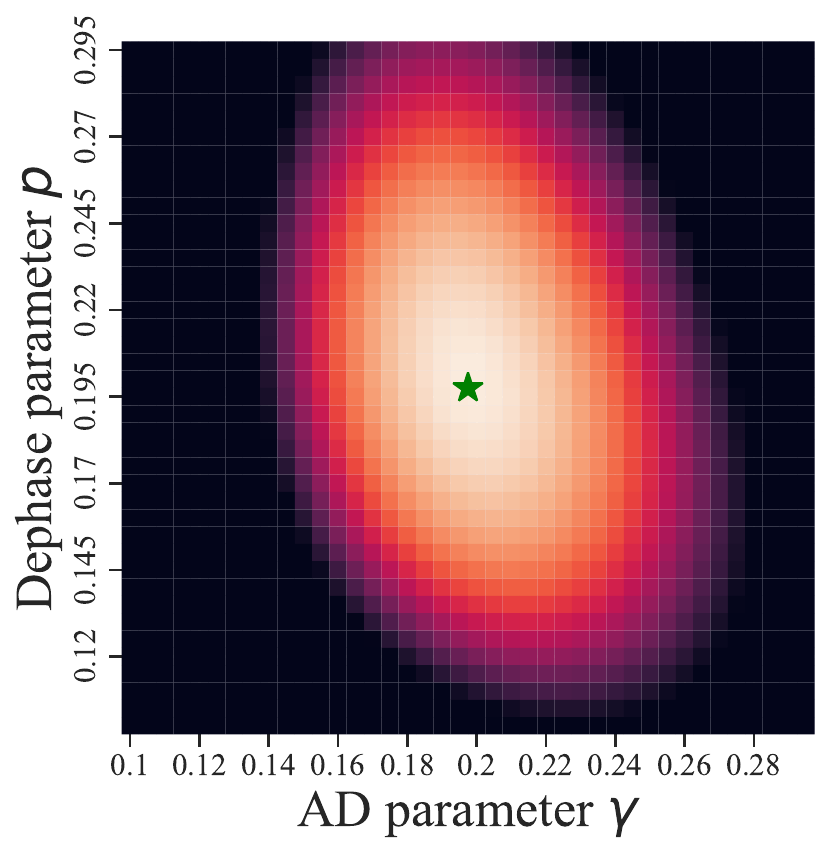}
        \put(-120,120){{\textsf{(a)}}}
    \end{minipage}
    \begin{minipage}[b]{0.48\linewidth}
        \centering
        \includegraphics[width=\hsize]{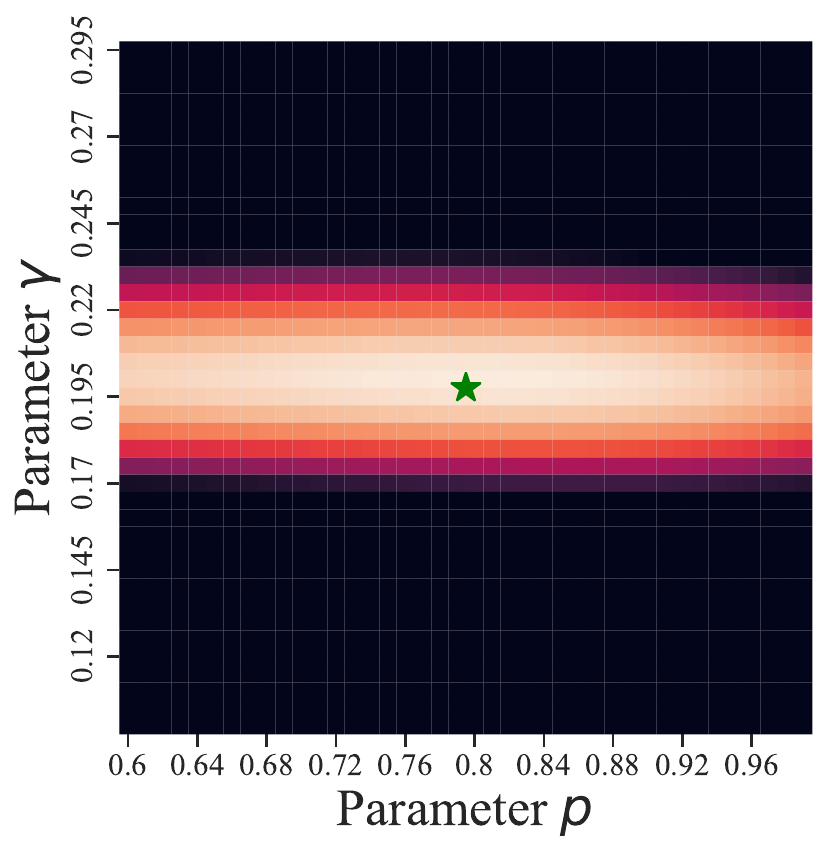}
        \put(-120,120){{\textsf{(b)}}}
    \end{minipage}
    \caption{
        Heatmaps of likelihood functions $p(\bm{m}|\bm{\alpha})$ for (a) AD$+$dephase and (b) GAD.
        The code size is $5\times 5$, and the number of syndrome measurement sets $n$ is $10^4$.
        The green stars represent the true model parameters.
    }\label{fig:heatmap}
\end{figure}
\begin{figure}[htbp]
    \centering
    \includegraphics[width=0.95\hsize]{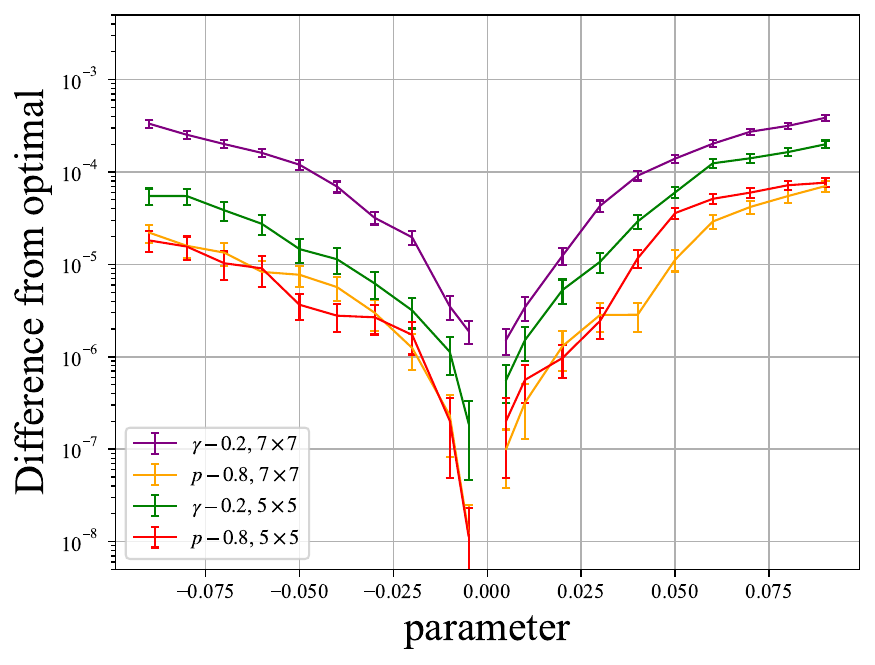}
    \caption{
        Dependence of decoder performance on the noise parameters of the GAD model.
        The decoder performance is evaluated by the expected diamond norm distance $||\mathcal{D} \circ \mathcal{M} \circ \mathcal{E} - I||_{\diamond}$, computed from $10^5$ samples of syndrome measurements, and is plotted as the deviation from the optimal value.
        Each plot corresponds to a different combination of code size and noise parameter: purple for $\gamma$ with a $7\times7$ code, orange for $p$ with a $7\times7$ code, green for $\gamma$ with a $5\times5$ code, and red for $p$ with a $5\times5$ code.
    }\label{fig:DecoderDependence}
\end{figure}
\begin{figure}[htbp]
    \centering
    \includegraphics[width=0.7\hsize]{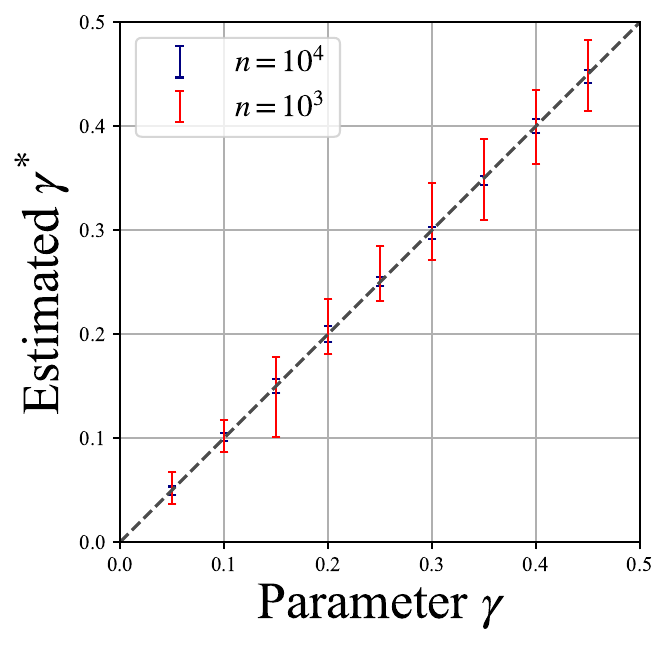}
    \caption{
        Results of estimation for nonuniform noise model.
        The code size is $3\times 3$, and the noise model is the amplitude damping noise model with $\gamma_{ij}=0.05(3i+j+1)$ for qubit at $(i,j)$.
    }\label{fig:NUAD}
\end{figure}
\begin{figure}[htbp]
    \centering
    \includegraphics[width=0.7\hsize]{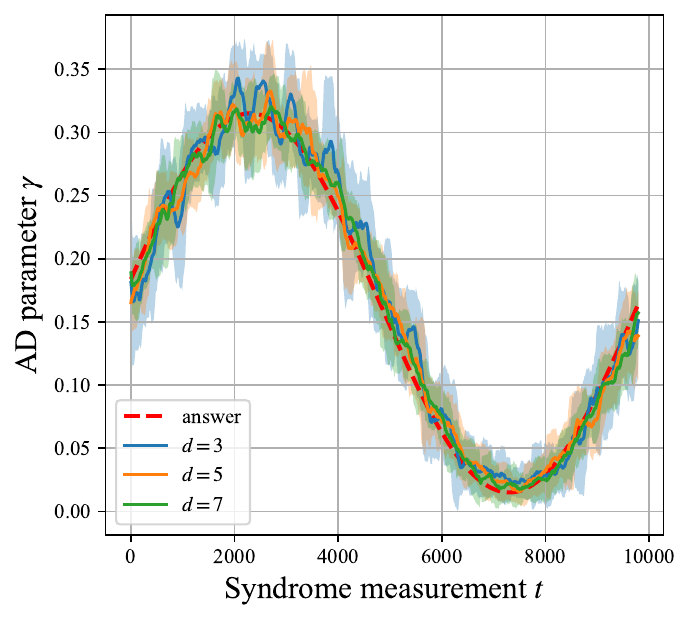}
    \caption{
        Results of estimation using SMC of sin-like time-varying noise model parameter of amplitude damping.
        The color range represents fluctuation before the smoothing process.
        The number of particles is $N=1280$, the resampling interval is $T=10$, and the smoothing time is $t_s=20$.
        The contraction of the TN is taken exactly.
    }\label{fig:SMC_sin}
\end{figure}
The results for the AD$+$dephasing noise model are shown in \reffig{fig:ADdephase}, where the number of syndrome measurement sets is $n = 10^2$.
We observe that the parameter estimation was successful.
As expected, the error bars are larger compared to the one-parameter estimation case.
To achieve a similar level of accuracy in multi-parameter estimations, a larger number of syndrome measurement results is required.

Next, we attempted parameter estimation for the GAD noise model, but the estimation failed, as shown in \reffig{fig:GAD}.
While the parameter corresponding to the strength of amplitude damping can be estimated, the parameter representing the directionality of the damping transitions—i.e., whether the transition occurs from $\ket{0}$ to $\ket{1}$ or vice versa—cannot be accurately determined.
The underlying reason is made clear by the heatmap of the likelihood function in \reffig{fig:heatmap}.
In the AD$+$dephasing case, the likelihood function exhibits a sharp peak, allowing for precise parameter estimation.
In contrast, in the GAD case, the likelihood exhibits much weaker dependence on $p$ compared to $\gamma$.
This disparity hinders reliable estimation of $p$, even when the prior distribution is optimized.
This estimability issue lies in the dependence of the noise model parameters on the noise map.
The explicit expressions of the one-qubit noisy state $\rho$ after noise map $\mathcal{E}_{dephase}\circ\mathcal{E}_{AD}(\rho)$ and $\mathcal{E}_{GAD}(\rho)$ are
\begin{align}
   \mathcal{E}_{dephase}\circ\mathcal{E}_{AD}(\rho) = \begin{pmatrix}
        \rho_{00} +\gamma\rho_{11} & \rho_{01}\sqrt{1-\gamma}\sqrt{1-p} \\
        \rho_{10}\sqrt{1-\gamma}\sqrt{1-p} & \rho_{11}(1-\gamma)
    \end{pmatrix}
\end{align}
and
\begin{align}
    &\mathcal{E}_{GAD}(\rho) = \\
    &\begin{pmatrix}
        (1-\gamma(1-p))\rho_{00} +\gamma p\rho_{11} & \rho_{01}\sqrt{1-\gamma} \\
        \rho_{10}\sqrt{1-\gamma} & (1-\gamma p)\rho_{11} +\gamma(1-p)\rho_{00}
    \end{pmatrix}.
\end{align}
The code-capacity noise model can be expressed as syndrome extraction process followed by the noise map $\mathcal{E}$ and the likelihood function $p_{\mathcal{E}}(\bm{m})$ and is calculated like
\begin{align}
    p(\bm{m}|\bm{\alpha}) &= \Tr\left[\prod_{i}\frac{1}{2}\left(I+m_{i}g_i\right)\mathcal{E}(\rho_n)\right]\\
    &= \frac{1}{2^{n-1}}\sum_{S_i\in\mathcal{S}}f(\bm{m},S_i)\Tr\left[S_i\mathcal{E}(\rho_n)\right],
\end{align}
where $n$ is the number of qubits, $\mathcal{S}$ is the stabilizer group for given code, and $f(\bm{m},S_i)$ is the function that returns the syndrome measurement result $\pm1$ for $S_i$ calculated from the results $\bm{m}$ for stabilizer generator.
From the above equations, we can evaluate the dependency of the likelihood function on the noise model parameters from $\Tr\left[S_i\mathcal{E}(\rho)\right]$.
To evaluate the dependency of the parameter in AD+dephase noise model and GAD, more clearly, we will focus onto the evaluations of the trace of one-qubit map $p_x=\Tr(X\mathcal{E}(\rho))$, $p_y=\Tr(Y\mathcal{E}(\rho))$, and $p_z=\Tr(Z\mathcal{E}(\rho))$ for rough and intuitive understanding.
For the AD$+$dephase noise model, we can see that the trace of the one-qubit map is
\begin{align}
    p_x &= 2\sqrt{1-\gamma}\sqrt{1-p}\Re(\rho_{10}),\\
    p_y &= 2\sqrt{1-\gamma}\sqrt{1-p}\Im(\rho_{10}),\\
    p_z &= \rho_{00}+(2\gamma - 1)\rho_{11}.
\end{align}
On the other hand, for the GAD noise model, we can see that the trace of the one-qubit map is
\begin{align}
    p_x &= 2\sqrt{1-\gamma}\Re(\rho_{10}),\\
    p_y &= 2\sqrt{1-\gamma}\Im(\rho_{10}),\\
    p_z &= (1-2\gamma(1-p))\rho_{00}+(2\gamma p-1)\rho_{11}.
\end{align}
When the $\gamma$ value is small, we can see that the dependence of $p$ is stronger for the AD$+$dephase noise model than for the GAD noise model.
For the AD$+$dephase noise model, the dependence of $p$ is not weak because of $\partial p_x/\partial p=-\sqrt{1-\gamma}\Re(\rho_{10})/\sqrt{1-p}$ and $\partial p_y/\partial p=-\sqrt{1-\gamma}\Im(\rho_{10})/\sqrt{1-p}$.
On the other hand, for the GAD noise model, the dependence of $p$ is weak because of $\partial p_x/\partial p=\partial p_y/\partial p=0$ and $\partial p_z/\partial p=2\gamma(\rho_{00}+\rho_{11})$.
It implies that the noise model parameter $p$ in the GAD noise model is more difficult to estimate than the one in the AD$+$dephase noise model, which is consistent with the results in \reffig{fig:heatmap}.
In general, the estimability of noise model parameters is difficult to evaluate and it will remain as an open problem.

Finally, we discuss the impact of the GAD noise model parameters on decoding performance, as shown in \reffig{fig:DecoderDependence}.
The results indicate that the dependence on $p$ is significantly weaker than that on $\gamma$, even for small code sizes.
Regarding code size, the influence of $\gamma$ becomes more pronounced as the code size increases, whereas the effect of $p$ remains nearly constant regardless of code size.
We therefore conclude that, for practical code sizes such as $d=20$, the decoder performance is much more sensitive to variations in $\gamma$ than in $p$.
This behavior is consistent with the fact that both the proposed parameter estimation and the maximum-likelihood decoding are governed by the structure of the likelihood function, which itself exhibits stronger dependence on $\gamma$.
To summarize, certain noise model parameters are inherently difficult to estimate; however, their impact on decoding performance is relatively minor and can often be neglected in practical applications.

\subsubsection{Nonuniform case}
Next, we show the results for the nonuniform noise model cases, where the noise model parameter depends on the qubit position.
The code size is $3\times 3$, and the target noise model is the nonuniform amplitude noise model with $\gamma_{ij}=0.05(3i+j+1)$ for qubit at $(i,j)$.
The results are put together in \reffig{fig:NUAD}.
We can see the estimation of non-uniform noise parameters was successful. However, more syndrome measurement results are needed to achieve a similar accuracy as the uniform case, \reffig{fig:one_para}(a).

\subsection{Results of sequential Monte Carlo for time-varying noise models}
In this section, we show the estimation results by using the TN simulator and SMC (\refalg{alg:qee SMC}) for time-varying noise models, more realistic situations than the previous section.
\begin{figure}[tbp]
    \begin{minipage}[t]{0.80\linewidth}
        \begin{overpic}[keepaspectratio,width=\hsize]{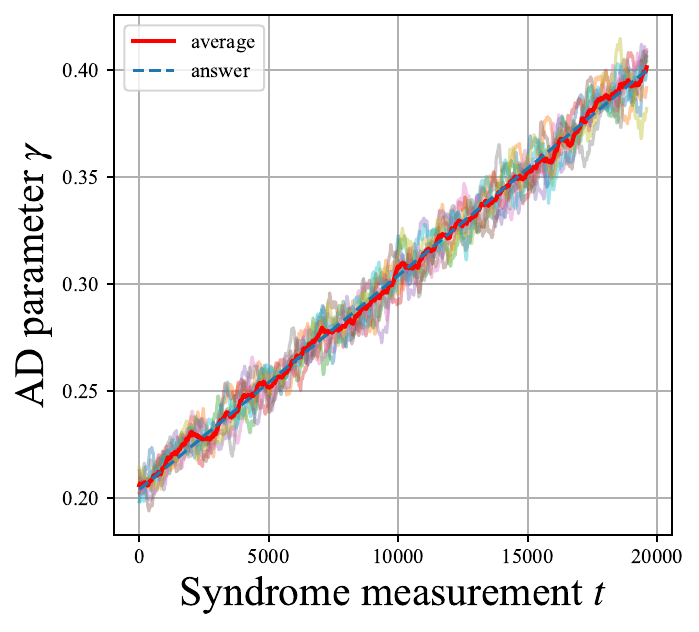}
        \end{overpic}
        \put(-180,180){{\textsf{(a)}}}
    \end{minipage}
    \begin{minipage}[t]{0.80\linewidth}
        \begin{overpic}[keepaspectratio,width=\hsize]{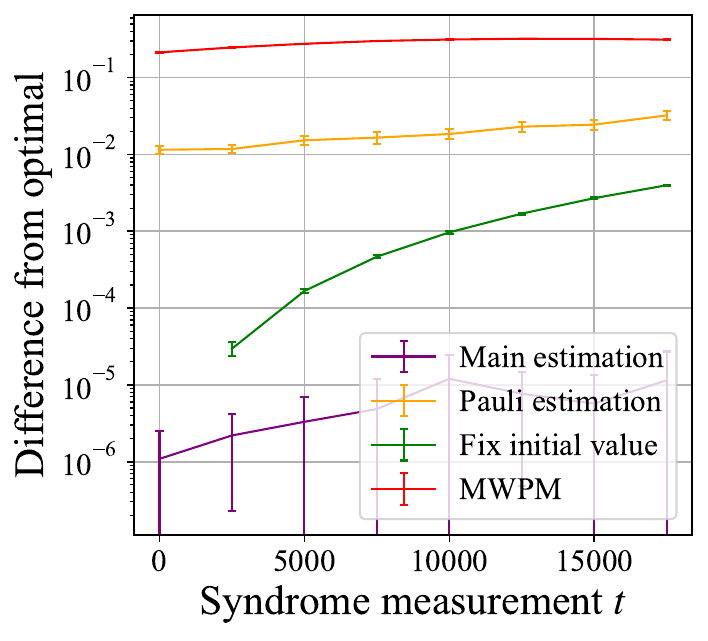}
        \end{overpic}
        \put(-180,180){{\textsf{(b)}}}
    \end{minipage}
    \caption{
        Results of estimation for line-like time-varying noise model and decode performance for code size $5\times 5$, the number of particles $N=1280$, resampling interval $T=10$, smoothing time $t_s=40$.
        (a) Estimation results of the time-varying noise model parameter for 10 different trials. Each translucent line represents the estimation result of each trial.
        (b) Difference from optimal performance of decoder performance using accurate knowledge of noise model for four patterns: (purple) TN decoder using estimation result, (yellow) TN decoder using estimation result assuming Pauli noise models, (green) TN decoder fixing the noise model parameter to initial value $\gamma=0.2$, (red) simple MWPM decoder without noise model information.
        The smaller results are better.
        Each point is calculated by the expected values of diamond distance $||\mathcal{D}\circ\mathcal{M}\circ\mathcal{E} - I||_{\diamond}$ over 10 different trials, calculated from $10^5$ samples.
        The initial point of the green line is omitted because the performance is perfectly matched with the optimal result.
        The performance TN simulations do not use any approximation.
    }\label{fig:SMC_line}
\end{figure}
\begin{figure}[tbp]
    \centering
    \includegraphics[width=0.7\hsize]{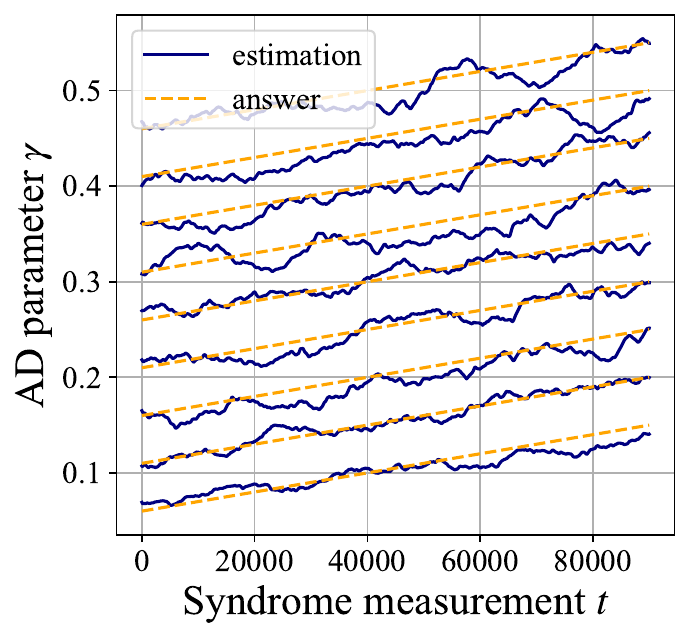}
    \caption{
    Results of estimation for time-varying nonuniform noise model case.
    Each line denotes the results for each qubit.
    The code size is $3\times 3$, and the noise models are the amplitude damping noise models whose parameters are $\gamma_{ij}(t) = a(3i+j+1)+bt$ with $a=0.05$ and $b=10^{-6}$ for qubit at $(i,j)$.
    TN simulations do not use any approximation.
    }\label{fig:sequential NUAD}
\end{figure}

\subsubsection{Amplitude damping noise model and the effect for decoder performance}
First, we show the results for the time-varying amplitude damping noise models.
We choose noise model parameters as a sin-like function as $\gamma(t) = a(b+\sin(2\pi \omega t))$ with $a=0.15$, $b=1.1$, and $\omega=10^{-4}$.
The posterior distribution is a random walk distribution in \refalg{alg:qee SMC}.
The results are presented in \reffig{fig:SMC_sin}.
We can see that the estimated parameters track the time-varying noise model parameter reasonably well.
As with previous estimation results, the larger the code size is, the more accurate the estimations are.

Next, we consider the line-like time-varying noise model and discuss the influence of estimations on decoding performance.
The noise model parameter is taken as $\gamma(t)=a+bt$ with $a=0.2$ and $b = 10^{-5}$.
The estimation results are shown in \reffig{fig:SMC_line}(a) and comparisons of decoding performance under various situations are shown in \reffig{fig:SMC_line}(b).

We can see that the estimations can track the time-varying noise parameter very well in \reffig{fig:SMC_line}(a).
In \reffig{fig:SMC_line}(b), using the estimation results, the TN decoder can outperform the TN decoder without the update of noise model parameters, which is the green line, and the simple MWPM decoder without noise model information, which is the red line.
To claim the need for estimation beyond the Pauli noise models, we also perform our estimation assuming the Pauli noise models, which is the yellow line.
We can see that the decoding performance of the main method is better than that of the Pauli noise models.
Thus, we conclude that selecting the appropriate noise model highly affects the decoding performance of the TN decoder and noise model estimation for general noise models is necessary.

\subsubsection{Nonuniform case}
Our SMC estimator method is expected to work well for estimating multi-parameter noise models as the MCMC estimator.
In this section, we present the results for the nonuniform line-like noise models.
The noise model is set as $\gamma_{ij}(t) = a(3i+j+1)+bt$ with $a=0.05$ and $b=10^{-6}$ for qubit at $(i,j)$.
The results are shown in \reffig{fig:sequential NUAD}.
The results demonstrate that the noise model is successfully estimated, even in the case of a multi-parameter nonuniform noise model.

%% file: conclusion.tex
\section{Summary and future issues}\label{sec:conclusion}
In this paper, we propose novel noise model estimation methods based on syndrome measurement results using the TN and Monte Carlo methods.
As seen in Sec.~\ref{sec:result}, the proposed methods work well for various noise models.
The TN simulator of the surface code can efficiently simulate arbitrary noise models for physical qubits.
Thus, the proposed estimation method can be applied to a broader range of noise models than just the Pauli noise models.
This work is expected to broaden the study of decoding algorithms that utilize general noise models, thereby accelerating the realization of fault-tolerant quantum computing.

The proposed methods are based on Bayesian inference.
Thus, even when the assumed noise model differs from the actual noise model in the device, the estimation can still work, and its outcomes are evaluated probabilistically.
Through evaluating the uncertainty of the estimation, we may use it for noise model selection~\cite{David1994, Raftery1997}, which will contribute to the improvement of the decoding performance as demonstrated in \reffig{fig:SMC_line}.
It is one of the advantages of using Bayesian inference not found in point estimations.
While the proposed methods, in principle, can apply to arbitrary noise models, we found that some noise models, such as the generalized amplitude damping, cannot be estimated.
It is natural because the estimations are based on the syndrome measurement results and initial logical information that cannot give us complete information about the quantum process.
From the perspective of efficient maximum-likelihood surface code decoding, however, this may not pose a significant disadvantage. 
Since the estimability of each noise model parameter is determined by its contribution to the likelihood function in Bayesian inference, a reduced impact implies that the precise values of these parameters are not always critical for likelihood-based decoding calculations.

Several factors determine the estimation accuracy in the proposed noise model estimation schemes.
One is the statistical fluctuation in the posterior distribution due to the finite number of samples of the syndrome measurements.
The second is the statistical error resulting from the finite Monte Carlo steps in MCMC and the finite number of particles in SMC.
To reduce the statistical error, for the SMC methods, the choice of the proposal distribution is crucial.
In this paper, we use the random walk distribution, which is a simple choice.
If the proposed distributions catch the time-varying trend more, the estimations will be more accurate. 
For MCMC methods, we can use more sophisticated methods, such as the exchange Monte Carlo method~\cite{Swendsen1986,Hukushima1996}.
Such exchange Monte Carlo methods can also deal with the multimodal likelihood function which is appearing in our numerical results.
The last is the systematic bias from the low-rank tensor approximation.
We must appropriately optimize the hyperparameters to achieve accurate estimations.
However, we can eliminate the last one, the systematic bias in tensor approximation, by using the Monte Carlo sampling in tensor network contraction~\cite{Ferris2015,Todo2024}.
Our proposed estimation methods have already been developed by combining the TN and Monte Carlo techniques.
Thus, integrating the tensor network Monte Carlo sampling method into our framework must be straightforward.

As for the estimation time, Monte Carlo simulations generally take a long time.
It is possible to shorten the estimation time by parallelizing the likelihood calculation, distributing SMC particles to many computation nodes, and introducing efficient low-rank approximation in the TN method.
However, it may not be enough for practical QEC, as the decoding time should be shorter than the decoherence time.
There might be two options to overcome such difficulties.
One is to utilize a subset of the syndrome measurement results instead of the entire dataset, effectively thinning out the data.
Another is to restrict our simulation code size and divide the results of syndrome measurements.
If the noise models can be estimated accurately for a smaller code, e.g., $3\times 3$, dividing the measurement results can accelerate the TN simulation and improve the performance by using parallelization.

It should also be noted that our evaluated noise models are code-capacity noise models with Markovian behavior.
However, in practical scenarios, evaluating syndrome measurement errors within phenomenological or circuit-level noise models and the non-Markovian noise model are crucial.
To address these issues, our future work will focus on extending the proposed methods to handle such noise models: TN methods and Monte Carlo methods for phenomenological and circuit-level noise models, or non-Markovian noise models.

%% file: ack.tex
\begin{acknowledgments}
This work was supported by the Center of Innovation for Sustainable Quantum AI, JST Grant Number JPMJPF2221, and JSPS KAKENHI Grant Numbers 20H01824 and 24K00543.
TK acknowledges the support of JST SPRING Grant Number JPMJSP2108.
The computation in this work has been partially done using the AI cluster system at the Institute for Physics of Intelligence, the University of Tokyo, and the facilities of the Supercomputer Center, the Institute for Solid State Physics, the University of Tokyo.
\end{acknowledgments}